# Energy-Efficient Design of MIMO Heterogeneous Networks with Wireless Backhaul

Howard H. Yang, *Student Member, IEEE,* Giovanni Geraci, *Member, IEEE,* and Tony Q. S. Quek, *Senior Member, IEEE*

***Abstract*—As future networks aim to meet the ever-increasing requirements of high data rate applications, dense and heterogeneous networks (HetNets) will be deployed to provide better coverage and throughput. Besides the important implications for energy consumption, the trend towards densification calls for more and more wireless links to forward a massive backhaul traffic into the core network. It is critically important to take into account the presence of a wireless backhaul for the energy-efficient design of HetNets. In this paper, we provide a general framework to analyze the energy efficiency of a two-tier MIMO heterogeneous network with wireless backhaul in the presence of both uplink and downlink transmissions. We find that under spatial multiplexing the energy efficiency of a HetNet is sensitive to the network load, and it should be taken into account when controlling the number of users served by each base station. We show that a two-tier HetNet with wireless backhaul can be significantly more energy efficient than a one-tier cellular network. However, this requires the bandwidth division between radio access links and wireless backhaul to be optimally designed according to the load conditions.**

## I. Introduction

In order to meet the exponentially growing mobile data demand, the next generation of wireless communication systems targets a thousand-fold capacity improvement, and the prospective increase in energy consumption poses urgent environmental and economic challenges [3], [4]. Green communications have become an inevitable necessity, and much effort is being made both in industry and academia to develop new architectures that can reduce the energy per bit from current levels, thus ensuring the sustainability of future wireless networks [5]–[9].

### *A. Background and Motivation*

Since the current growth rate of wireless data exceeds both spectral efficiency advances and availability of new wireless spectrum, a trend towards densification and heterogeneity is essential to respond adequately to the continued surge in mobile data traffic [10]–[12]. To this end, heterogeneous networks (HetNets) can provide higher coverage and throughput by overlaying macro cells with a large number of small cells and access points, thus offloading traffic and reducing the distance between transmitter and receiver [13], [14]. When small cells are densely deployed, forwarding a massive cellular traffic to the backbone network becomes a key problem, and a wireless backhaul is regarded as the only practical solution for outdoor scenarios where wired links are not available [15]–[19]. However, the power consumption incurred on the wireless backhaul links, together with the power consumed by the multitude of access points deployed, becomes a crucial issue, and an energy-efficient design is necessary to ensure the viability of future wireless HetNets [20].

Various approaches have been investigated to improve the energy efficiency of heterogeneous networks. Cell size, deployment density, and number of antennas were optimized to minimize the power consumption of small cells [21], [22]. Cognitive sensing and sleep mode strategies were also proposed to turn off inactive access points and enhance the energy efficiency [23], [24]. A further energy efficiency gain was shown to be attainable by serving users that experience better channel conditions, and by dynamically assigning users to different tiers of the network [25], [26]. Although various studies have been conducted on the energy efficiency of HetNets, the impact of a wireless backhaul has typically been neglected. On the other hand, the power consumption of backhauling operations at small cell access points (SAPs) might be comparable to the amount of power necessary to operate macro base stations (MBSs) [27]–[29]. Moreover, since it is responsible to aggregate traffic from SAPs towards MBSs, the backhaul may significantly affect the rates and therefore the energy efficiency of the entire network. With a potential evolution towards dense infrastructures, where many small access points are expected to be used, it is of critical importance to take into account the presence of a wireless backhaul for the energy-efficient design of heterogeneous networks.

### *B. Approach and Main Outcomes*

The main goal of this paper is to study the energy efficiency of heterogeneous networks with wireless backhaul. We consider a two-tier HetNet which consists of MBSs and SAPs, where SAPs are connected to MBSs via a multiple-input-multiple-output (MIMO) wireless backhaul that uses a fraction

Manuscript received September 15, 2015; revised January 24, 2016; accepted March 17, 2016. The associate editor coordinating the review of this manuscript and approving it for publication was Dr. Mohammad Reza Nakhai.

This work was supported in part by the A*STAR SERC under Grant 1224104048, and the MOE ARF Tier 2 under Grant MOE2014-T2-2-002, and the SUTD-ZJU Research Collaboration under Grant SUTD-ZJU/RES/01/2014. This work was presented in part at 2016 IEEE International Conference on Acoustics, Speech and Signal Processing [1], and will be presented in part to 2016 IEEE International Conference on Communications [2].

H. H. Yang, and T. Q. S. Quek are with the Singapore University of Technology and Design, Singapore (hao_yang@mymail.sutd.edu.sg, tonyquek@sutd.edu.sg). G. Geraci is with Bell Labs Nokia, Ireland ( giovanni.geraci@nokia.com).

of the total available bandwidth. We undertake an analytical approach to derive data rates and power consumption for the entire network in the presence of both uplink (UL) and downlink (DL) transmissions and spatial multiplexing. This is a practical scenario that has not yet been addressed. In this paper, we model the spatial locations of MBSs, SAPs, and user equipments (UEs) as independent homogeneous Poisson point processes (PPPs), and analyze the energy efficiency by combining tools from stochastic geometry and random matrix theory. Stochastic geometry is a powerful tool to analyze the interference in large HetNets with a random topology [30], whereas random matrix theory enables a deterministic abstraction of the physical layer, for a fixed network topology [31]. Our analysis is general and encompasses all the key features of a heterogeneous network, i.e., interference, load, deployment strategy, and capability of the wireless infrastructure components. With the developed framework, we can explicitly characterize the power consumption of the HetNet due to signal processing operations in macro cells, small cells, and wireless backhaul, as well as the rates and ultimately the energy efficiency of the whole network. Our main contributions are summarized below.

- We provide a general toolset to analyze the energy efficiency of a two-tier MIMO heterogeneous network with wireless backhaul. Our model accounts for both UL and DL transmissions and spatial multiplexing, for the bandwidth and power allocated between macro cells, small cells, and backhaul, and for the infrastructure deployment strategy.
- We combine tools from stochastic geometry and random matrix theory to derive the uplink and downlink rates of macro cells, small cells, and wireless backhaul. The resulting analysis is tractable and captures the effects of multiantenna transmission, fading, shadowing, and random network topology.
- Using the developed framework, we find that the energy efficiency of a HetNet is sensitive to the load conditions of the network, thus establishing the importance of scheduling the right number of UEs per base station when spatial multiplexing is employed. Moreover, by comparing the energy efficiency under different deployment scenarios, we find that such property does not depend on the infrastructure.
- We show that if the wireless backhaul is not allocated sufficient resources, then the energy efficiency of a two-tier HetNet with wireless backhaul can be worse than that of a one-tier cellular network. However, the two-tier HetNet can achieve a significant energy efficiency gain if the backhaul bandwidth is optimally allocated according to the load conditions of the network.

The remainder of the paper is organized as follows. The system model is introduced in Section II. In Section III, we detail the power consumption of a heterogeneous network with wireless backhaul. In Section IV, we analyze the data rates and the energy efficiency, and we provide simulations that confirm the accuracy of our analysis. Numerical results are shown in Section V to give insights into the energy-efficient design of a HetNet with wireless backhaul. The paper is concluded in Section VI.

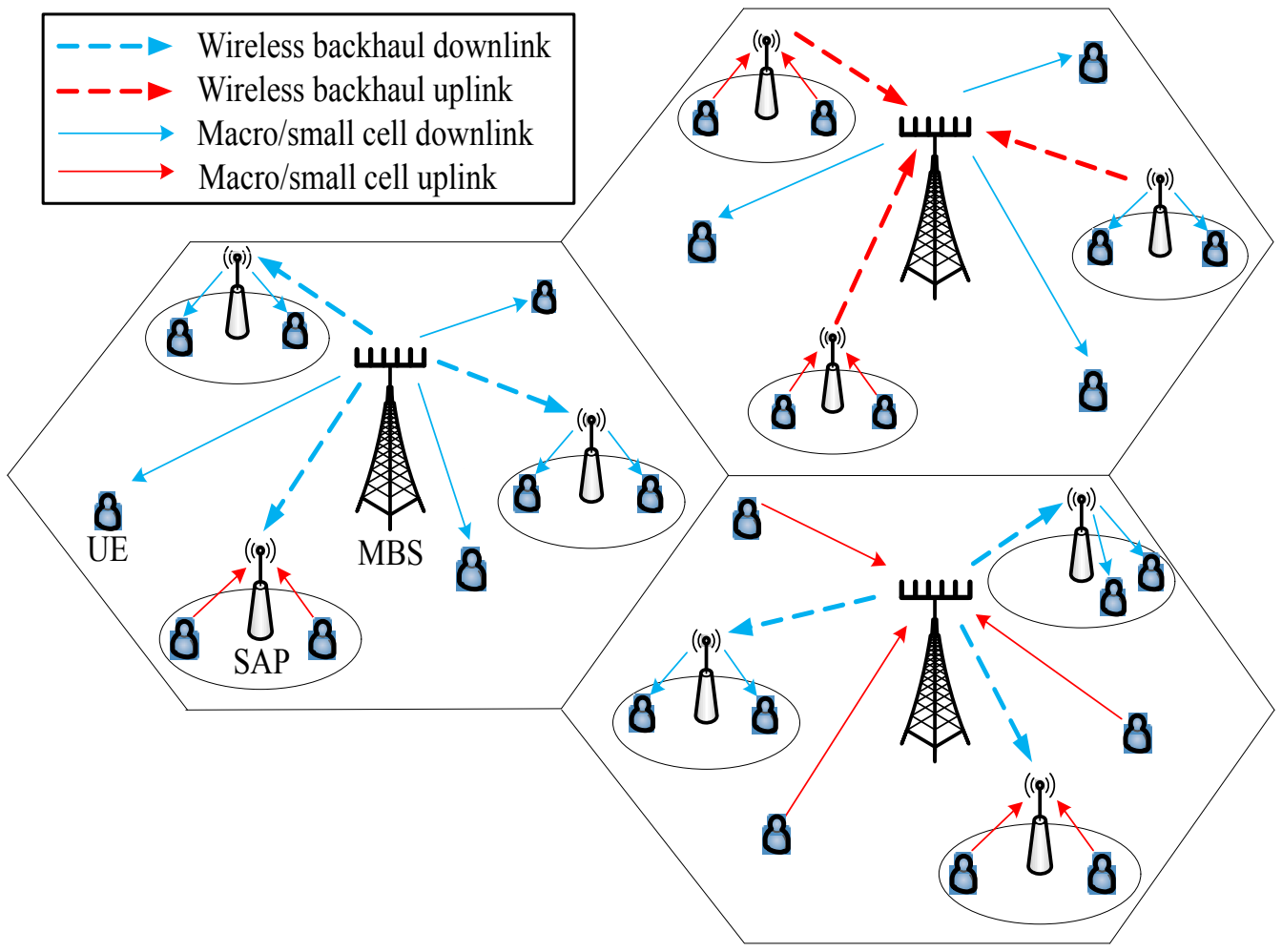

Fig. 1. Illustration of a two-tier heterogeneous network with wireless backhaul.

## II. System Model

### A. Topology and Channel

We study a two-tier heterogeneous network which consists of MBSs, SAPs, and UEs, as depicted in Figure 1. The spatial locations of MBSs, SAPs, and UEs follow independent PPPs $\Phi_{\mathrm{m}}$, $\Phi_{\mathrm{s}}$, and $\Phi_{\mathrm{u}}$, with spatial densities $\lambda_{\mathrm{m}}$, $\lambda_{\mathrm{s}}$, and $\lambda_{\mathrm{u}}$, respectively[1]. All MBSs, SAPs, and UEs are equipped with $M_{\mathrm{m}}$, $M_{\mathrm{s}}$, and 1 antennas, respectively, each UE associates with the base station that provides the largest average received power, and each SAP associates with the closest MBS. The links between MBSs and UEs, SAPs and UEs, and MBSs and SAPs are referred to as *macro cell links*, *small cell links*, and *backhaul links*, respectively. In light of its higher spectral efficiency [34], we consider spatial multiplexing where each MBS and each SAP simultaneously serve $K_{\mathrm{m}}$ and $K_{\mathrm{s}}$ UEs, respectively. In practice, due to a finite number of antennas, MBSs and SAPs use traffic scheduling to limit the number of UEs served to $K_{\mathrm{m}} \leq M_{\mathrm{m}}$ and $K_{\mathrm{s}} \leq M_{\mathrm{s}}$ [35]. Similarly, each MBS limits to $K_{\mathrm{b}}$ the number of SAPs served on the backhaul, with $K_{\mathrm{b}} M_{\mathrm{s}} \leq M_{\mathrm{m}}$. The MIMO dimensionality ratio for linear processing on macro cells, small cells, and backhaul is denoted by $\beta_{\mathrm{m}} = \frac{K_{\mathrm{m}}}{M_{\mathrm{m}}}$, $\beta_{\mathrm{s}} = \frac{K_{\mathrm{s}}}{M_{\mathrm{s}}}$, and $\beta_{\mathrm{b}} = \frac{K_{\mathrm{b}} M_{\mathrm{s}}}{M_{\mathrm{m}}}$, respectively. As the MIMO dimensionality ratio at a base station reveals the number of active UEs within the cell coverage, for simplicity of notation, we refer to the MIMO dimensionality ratio as load whenever it does not cause ambiguity.

In this work, we consider a co-channel deployment of small cells with the macro cell tier, i.e., macro cells and small cells share the same frequency band for transmission[2]. In order to avoid severe interference which may degrade the performance of the network, we assume that the access and backhaul links share the same pool of radio resources through orthogonal division, i.e., the total available bandwidth is divided into two portions, where a fraction $\zeta_{\mathrm{b}}$ is used for the wireless backhaul, and the remaining $(1-\zeta_{\mathrm{b}})$ is shared by the radio access links (macro cells and small cells) [15], [17], [40], [41]. In order to adapt the radio resources to the variation of the DL/UL traffic demand, we assume that MBSs and SAPs operate in a dynamic time division duplex (TDD) mode [42], [43], where at every time slot, all MBSs and SAPs independently transmit in downlink with probabilities $\tau_{\mathrm{m}}$, $\tau_{\mathrm{s}}$, and $\tau_{\mathrm{b}}$ on the macro cell, small cell, and backhaul, respectively, and they transmit in uplink for the remaining time[3]. We model the channels between any pair of antennas in the network as independent, narrowband, and affected by three attenuation components, namely small-scale Rayleigh fading, shadowing $S_{\mathrm{D}}$ and $S_{\mathrm{B}}$ for data link and backhaul link, respectively, and large-scale path loss, where $\alpha$ is the path loss exponent and the shadowing satisfies $\mathbb{E}[S_{\mathrm{D}}^{\frac{2}{\alpha}}] < \infty$ and $\mathbb{E}[S_{\mathrm{B}}^{\frac{2}{\alpha}}] < \infty$, and by thermal noise with variance $\sigma^2$. We finally assume that all MBSs and SAPs use a zero forcing (ZF) scheme for both transmission and reception, due to its practical simplicity [44][4].

[1] Note that a PPP can serve as a good model not only for the opportunistic deployment of small cell access points, but also for the planned deployment of macro cell base stations, as verified by both empirical evidence [32] and theoretical analysis [33].

### B. Energy Efficiency

We consider the power consumption due to transmission and signal processing operations performed on the entire network, therefore energy-efficiency tradeoffs will be such that savings at the MBSs and SAPs are not counteracted by increased consumption at the UEs, and vice versa [6], [49]. We can identify three main contributions to the power consumption of the heterogeneous network, namely the consumption on macro cells, small cells, and backhaul links. Consistent with previous work [49]–[52], we account for the power consumption due to transmission, encoding, decoding, and analog circuits. A detailed model for the power consumption of the HetNet will be given in Section III.

Let $\mathcal{P}[\frac{\mathrm{W}}{\mathrm{m}^2}]$ be the total power consumption per area, which includes the power consumed on all links. We denote by $\mathcal{R}[\frac{\mathrm{bit}}{\mathrm{m}^2}]$ the sum rate per unit area of the network, i.e., the total number of bits per second successfully transmitted per square meter. The energy efficiency $\eta = \frac{\mathcal{R}}{\mathcal{P}}$ is then defined as the number of bits successfully transmitted per joule of energy spent [49], [53]. For the sake of clarity, the main notations used in this paper are summarized in Table I.

[2]Many frequency planning possibilities exist for MBSs and SAPs, where the optimal solution is traffic load dependent. Though a non-co-channel allocation is justified for highly dense scenarios [36]–[38], in some cases a co-channel deployment may be preferred from an operator's perspective, since MBSs and SAPs can share the same spectrum thus improving the spectral utilization ratio [39].

[3]We note that different SAPs and MBSs may have different uplink/downlink resource partitions for their associated UEs. Since the aggregate interference is affected by the average value of such partitions, we assume fixed and uniform uplink/downlink partitions.

[4]Note that the results involving the machinery of random matrix theory can be adjusted to account for different transmit precoders and receive filters, imperfect channel state information, and antenna correlation [45]–[48].

TABLE I
NOTATION SUMMARY

| Notation | Definition |
|---|---|
| $\mathcal{P}$; $\mathcal{R}$; $\eta$ | Power per area; rate per area; energy efficiency |
| $R_{\mathrm{m}}^{\mathrm{DL}}$; $R_{\mathrm{s}}^{\mathrm{DL}}$; $R_{\mathrm{b}}^{\mathrm{DL}}$ | Downlink rate on macro cells, small cells, and backhaul |
| $R_{\mathrm{m}}^{\mathrm{UL}}$; $R_{\mathrm{s}}^{\mathrm{UL}}$; $R_{\mathrm{b}}^{\mathrm{UL}}$ | Uplink rate on macro cells, small cells, and backhaul |
| $P_{\mathrm{mt}}$; $P_{\mathrm{st}}$; $P_{\mathrm{ut}}$ | Transmit power for MBSs, SAPs, and UEs |
| $P_{\mathrm{mb}}$; $P_{\mathrm{sb}}$ | Backhaul transmit power for MBSs and SAPs |
| $P_{\mathrm{mc}}$; $P_{\mathrm{sc}}$ | Analog circuit power consumption at macro cells and small cells |
| $P_{\mathrm{me}}$; $P_{\mathrm{se}}$; $P_{\mathrm{ue}}$ | Encoding power per bit on macro cells, small cells, and backhaul |
| $P_{\mathrm{md}}$; $P_{\mathrm{sd}}$; $P_{\mathrm{ud}}$ | Decoding power per bit on macro cells, small cells, and backhaul |
| $\Phi_{\mathrm{m}}$; $\Phi_{\mathrm{s}}$; $\Phi_{\mathrm{u}}$ | PPPs modeling locations of MBSs, SAPs, and UEs |
| $\lambda_{\mathrm{m}}$; $\lambda_{\mathrm{s}}$; $\lambda_{\mathrm{u}}$ | Spatial densities of MBSs, SAPs, and UEs |
| $A_{\mathrm{m}}$; $A_{\mathrm{s}}$ | Association probabilities for MBSs and SAPs |
| $M_{\mathrm{m}}$; $M_{\mathrm{s}}$ | Number of transmit antennas per MBS and SAP |
| $K_{\mathrm{m}}$; $K_{\mathrm{s}}$; $K_{\mathrm{b}}$ | UEs served per macro cell and small cell; SAPs per MBS on backhaul |
| $\beta_{\mathrm{m}}$; $\beta_{\mathrm{s}}$; $\beta_{\mathrm{b}}$ | Load on macro cells, small cells, and backhaul |
| $\tau_{\mathrm{m}}$; $\tau_{\mathrm{s}}$; $\tau_{\mathrm{b}}$ | Fraction of time in DL for macro cells, small cells, and backhaul |
| $\zeta_{\mathrm{b}}$; $\alpha$ | Fraction of bandwidth for backhaul; path loss exponent |
| $S_{\mathrm{D}}$; $S_{\mathrm{B}}$ | Shadowing on radio access link and wireless backhaul |

## III. POWER CONSUMPTION

In this section, we model in detail the power consumption of the heterogeneous network with wireless backhaul.

Since each UE associates with the base station, i.e., MBS or SAP, that provides the largest average received power, the probability that a UE associates to a MBS or to a SAP can be respectively calculated as [54]

$$A_{\mathrm{m}} = \frac{\lambda_{\mathrm{m}} P_{\mathrm{mt}}^{\frac{2}{\alpha}}}{\lambda_{\mathrm{m}} P_{\mathrm{mt}}^{\frac{2}{\alpha}} + \lambda_{\mathrm{s}} P_{\mathrm{st}}^{\frac{2}{\alpha}}} \tag{1}$$

and

$$A_{\mathrm{s}} = \frac{\lambda_{\mathrm{s}} P_{\mathrm{st}}^{\frac{2}{\alpha}}}{\lambda_{\mathrm{m}} P_{\mathrm{mt}}^{\frac{2}{\alpha}} + \lambda_{\mathrm{s}} P_{\mathrm{st}}^{\frac{2}{\alpha}}}. \tag{2}$$

In the remainder of the paper, we make use of the following approximation.

**Assumption 1:** *We approximate the number of UEs, the number of SAPs associated to a MBS, and the number of UEs associated to a SAP by constant values $K_{\mathrm{m}}$, $K_{\mathrm{b}}$, and $K_{\mathrm{s}}$, respectively, which are upper bounds imposed by practical antenna limitations at MBSs and SAPs*[5].

The assumption above is motivated by the fact that the

[5]The number of base station antennas imposes a constraint on the maximum number of UEs scheduled for transmission. In fact, under linear precoding, the number of scheduled UEs should not exceed the number of antennas, in order for the achievable rate not to be significantly degraded [55]–[57].

$$\mathcal{P} = \lambda_{\rm m}\left[\tau_{\rm m}P_{\rm mt}+(1-\tau_{\rm m})\,K_{\rm m}P_{\rm ut}+P_{\rm mf}+P_{\rm ma}M_{\rm m}+P_{\rm ua}K_{\rm m}+\tau_{\rm m}K_{\rm m}\left(P_{\rm me}+P_{\rm ud}\right)R_{\rm m}^{\rm DL}+(1-\tau_{\rm m})\,K_{\rm m}\left(P_{\rm md}+P_{\rm ue}\right)R_{\rm m}^{\rm UL}\right]$$
$$+\lambda_{\rm s}\left[\tau_{\rm s}P_{\rm st}+(1-\tau_{\rm s})\,K_{\rm s}P_{\rm ut}+P_{\rm sf}+P_{\rm sa}M_{\rm s}+P_{\rm ua}K_{\rm s}+\tau_{\rm s}K_{\rm s}\left(P_{\rm se}+P_{\rm ud}\right)R_{\rm s}^{\rm DL}+(1-\tau_{\rm s})\,K_{\rm s}\left(P_{\rm sd}+P_{\rm ue}\right)R_{\rm s}^{\rm UL}\right]+\lambda_{\rm m}\left[\tau_{\rm b}P_{\rm mb}\right.$$
$$\left.+(1-\tau_{\rm b})\,K_{\rm b}P_{\rm sb}+P_{\rm ma}M_{\rm m}+K_{\rm b}M_{\rm s}P_{\rm sa}+\tau_{\rm b}K_{\rm b}K_{\rm s}\left(P_{\rm me}+P_{\rm sd}\right)R_{\rm b}^{\rm DL}+(1-\tau_{\rm b})\,K_{\rm b}K_{\rm s}\left(P_{\rm md}+P_{\rm se}\right)R_{\rm b}^{\rm UL}\right]. \quad (10)$$

---

number of UEs $N_{\rm m}$ served by a MBS has distribution [54]

$$\mathbb{P}(N_{\rm m}=n)=\frac{3.5^{3.5}\Gamma(n+3.5)\left(\frac{\lambda_{\rm m}}{A_{\rm m}\lambda_{\rm u}}\right)^{3.5}}{\Gamma(3.5)n!\,(1+3.5\lambda_{\rm m}/\lambda_{\rm u})^{n+3.5}} \quad (3)$$

where $\Gamma(\cdot)$ is the gamma function. Let $K_{\rm m}$ be a limit on the number of users that can be served by a MBS, the probability that a MBS serves less than $K_{\rm m}$ UEs is given by

$$\mathbb{P}\left(N_{\rm m}<K_{\rm m}\right)=\sum_{n=0}^{K_{\rm m}-1}\frac{3.5^{3.5}\Gamma(n+3.5)\left(\frac{\lambda_{\rm m}}{A_{\rm m}\lambda_{\rm u}}\right)^{3.5}}{\Gamma(3.5)n!\,(1+3.5\lambda_{\rm m}/\lambda_{\rm u})^{n+3.5}}$$
$$\leq\left(\frac{2\lambda_{\rm m}}{\lambda_{\rm u}}\right)^{3.5}\sum_{n=0}^{K_{\rm m}-1}\frac{\Gamma(n+3.5)}{n!}\frac{3.5^{3.5}}{\Gamma(3.5)} \quad (4)$$

which rapidly tends to zero as $\frac{\lambda_{\rm u}}{\lambda_{\rm m}}$ grows. This indicates that in a practical network with a high density of UEs, i.e., where $\lambda_{\rm u}\gg\lambda_{\rm m}$, each MBS serves $K_{\rm m}$ UEs with probability almost one. A similar approach can be used to show that $\mathbb{P}(N_{\rm s}<K_{\rm s})\approx 0$ and $\mathbb{P}(N_{\rm b}<K_{\rm b})\approx 0$ when $\lambda_{\rm u}\gg\lambda_{\rm m}$ and $\lambda_{\rm s}\gg\lambda_{\rm m}$, respectively, and therefore each SAP serves $K_{\rm s}$ UEs and each MBS serves $K_{\rm b}$ SAPs on the backhaul with probability almost one.

In the following, we use the power consumption model introduced in [49], which captures all the key contributions to the power consumption of signal processing operations. This model is flexible since the various power consumption values can be tuned according to different scenarios. We note that the results presented in this paper hold under more general conditions and apply to different power consumption models [58], [59].

Under the previous assumption, and by using the model in [49], we can write the power consumption on each macro cell link as follows

$$P_{\rm m}=\tau_{\rm m}P_{\rm mt}+(1-\tau_{\rm m})\,K_{\rm m}P_{\rm ut}+\tau_{\rm m}K_{\rm m}\left(P_{\rm me}+P_{\rm ud}\right)R_{\rm m}^{\rm DL}$$
$$+P_{\rm mc}+(1-\tau_{\rm m})\,K_{\rm m}\left(P_{\rm md}+P_{\rm ue}\right)R_{\rm m}^{\rm UL} \quad (5)$$

where $P_{\rm mt}$ and $P_{\rm ut}$ are the DL and UL transmit power from the MBS and the $K_{\rm m}$ UEs, respectively, $P_{\rm mc}$ is the analog circuit power consumption, $P_{\rm me}$ and $P_{\rm md}$ are encoding and decoding power per bit of information for MBS, while $P_{\rm ue}$ and $P_{\rm ud}$ are encoding and decoding power per bit of information for UE, and $R_{\rm m}^{\rm DL}$ and $R_{\rm m}^{\rm UL}$ denote the DL and UL rates for each MBS-UE pair. The analog circuit power can be modeled as [49]

$$P_{\rm mc}=P_{\rm mf}+P_{\rm ma}M_{\rm m}+P_{\rm ua}K_{\rm m} \quad (6)$$

where $P_{\rm mf}$ is a fixed power accounting for control signals, baseband processor, local oscillator at MBS, cooling system, etc., $P_{\rm ma}$ is the power required to run each circuit component attached to the MBS antennas, such as converter, mixer, and filters, $P_{\rm ua}$ is the power consumed by circuits to run a single-antenna UE. Under this model, the total power consumption on the macro cell can be written as

$$P_{\rm m}=\tau_{\rm m}P_{\rm mt}+(1-\tau_{\rm m})\,K_{\rm m}P_{\rm ut}+\tau_{\rm m}K_{\rm m}(P_{\rm me}+P_{\rm ud})R_{\rm m}^{\rm DL}$$
$$+P_{\rm mf}+P_{\rm ma}M_{\rm m}+P_{\rm ua}K_{\rm m}+(1-\tau_{\rm m})K_{\rm m}(P_{\rm md}+P_{\rm ue})R_{\rm m}^{\rm UL}. \quad (7)$$

Through a similar approach, the power consumption on each small cell and backhaul link can be written as

$$P_{\rm s}=\tau_{\rm s}P_{\rm st}+(1-\tau_{\rm s})\,K_{\rm s}P_{\rm ut}+P_{\rm sf}+\tau_{\rm s}K_{\rm s}\left(P_{\rm se}+P_{\rm ud}\right)R_{\rm s}^{\rm DL}$$
$$+P_{\rm sa}M_{\rm s}+P_{\rm ua}K_{\rm s}+(1-\tau_{\rm s})\,K_{\rm s}\left(P_{\rm sd}+P_{\rm ue}\right)R_{\rm s}^{\rm UL} \quad (8)$$

and

$$P_{\rm b}=\tau_{\rm b}P_{\rm mb}+(1-\tau_{\rm b})\,K_{\rm b}P_{\rm sb}+\tau_{\rm b}K_{\rm b}K_{\rm s}\left(P_{\rm me}+P_{\rm sd}\right)R_{\rm b}^{\rm DL}$$
$$+P_{\rm ma}M_{\rm m}+K_{\rm b}M_{\rm s}P_{\rm sa}+(1-\tau_{\rm b})\,K_{\rm b}K_{\rm s}\left(P_{\rm md}+P_{\rm se}\right)R_{\rm b}^{\rm UL}, \quad (9)$$

respectively, the analog circuit power consumption in (9) accounts for power spent on out of band SAPs. In the above equations, $P_{\rm st}$ is the transmit power on a small cell, $P_{\rm mb}$ and $P_{\rm sb}$ are the powers transmitted by MBSs and SAPs on the backhaul, and $P_{\rm sf}$ and $P_{\rm sa}$ are the small-cell equivalents of $P_{\rm mf}$ and $P_{\rm ma}$. Moreover, $R_{\rm s}^{\rm DL}$ and $R_{\rm s}^{\rm UL}$ denote the DL and UL rates for each SAP-UE pair, and $R_{\rm b}^{\rm DL}$ and $R_{\rm b}^{\rm UL}$ denote the DL and UL rates for each wireless backhaul link.

We can now write the total power consumption of the heterogeneous network with wireless backhaul.

**Lemma 1:** *The power consumption per area in a heterogeneous network with wireless backhaul is given by* (10) *shown on the top of this page.*

*Proof:* Equation (10) follows from (7), (8), (9), and by noting that under Assumption 1 the average power consumption per area can be expressed as $\mathcal{P}=P_{\rm m}\lambda_{\rm m}+P_{\rm s}\lambda_{\rm s}+P_{\rm b}\lambda_{\rm m}$. □

## IV. Rates and Energy Efficiency

In this section, we analyze the data rates and the energy efficiency of a HetNet with wireless backhaul, and we provide simulations that verify the accuracy of our analysis. Particularly, we combine tools from stochastic geometry and random matrix theory to derive the uplink and downlink rates of macro cells, small cells, and wireless backhaul. The resulting analysis is tractable and captures the effects of multiantenna transmission, fading, shadowing, and random network topology. By using the framework developed in this section, we will show that the energy efficiency of a HetNet is sensitive to the load conditions of the network, irrespective of the infrastructure used, and that a two-tier HetNet can achieve a significant

$$R_{\rm m}^{\rm DL} = (1-\zeta_{\rm b})\int_0^\infty\int_0^\infty \frac{e^{-\sigma^2 z}}{z\ln 2}\left(1-e^{-z\nu_{\rm m}^{\rm D}}\right)\exp\left(-\frac{2\pi^2\tilde{\lambda}_{\rm u}P_{\rm ut}^{\delta}\mathbb{E}[S_{\rm D}^{\delta}]z^{\delta}}{\alpha\sin\left(\frac{2\pi}{\alpha}\right)}\right)$$
$$\times\exp\left(-\tau_{\rm m}a_{\rm m}\mathcal{C}_{\alpha,K_{\rm m}}(zP_{\rm mt},t)\left(\frac{zP_{\rm mt}}{K_{\rm m}}\right)^{\delta}-\tau_{\rm s}a_{\rm s}\mathcal{C}_{\alpha,K_{\rm s}}(zP_{\rm mt},t)\left(\frac{zP_{\rm st}}{K_{\rm s}}\right)^{\delta}\right)f_{L_{\rm m}}(t)dtdz \tag{11}$$

$$R_{\rm m}^{\rm UL} = (1-\zeta_{\rm b})\int_0^\infty\int_0^\infty \frac{e^{-\sigma^2 z}}{z\ln 2}\left(1-e^{-z\nu_{\rm m}^{\rm U}/t}\right)\exp\left\{-\tilde{\lambda}_{\rm u}\pi\mathbb{E}[S_{\rm D}^{\delta}]\int_0^\infty\frac{1-e^{-G_{\rm m}u}}{1+z^{-1}u^{\frac{1}{\delta}}/P_{\rm ut}}du\right.$$
$$\left.-\frac{\Gamma(1+\delta)\,\delta\pi^2 z^{\delta}}{\sin(\delta\pi)}\left[\frac{\tau_{\rm m}a_{\rm m}P_{\rm mt}^{\delta}\prod_{i=1}^{K_{\rm m}-1}(i+\delta)}{\Gamma(K_{\rm m})K_{\rm m}^{\delta}}+\frac{\tau_{\rm s}a_{\rm s}P_{\rm st}^{\delta}\prod_{i=1}^{K_{\rm s}-1}(i+\delta)}{\Gamma(K_{\rm s})K_{\rm s}^{\delta}}\right]\right\}f_{L_{\rm m}}(t)dtdz \tag{15}$$

$$R_{\rm s}^{\rm DL} = (1-\zeta_{\rm b})\int_0^\infty\int_0^\infty \frac{e^{-\sigma^2 z}}{z\ln 2}\left(1-\frac{1}{(1+zP_{\rm st}t^{-1}/K_{\rm s})^{\Delta_{\rm s}}}\right)\exp\left(-\frac{2\pi^2\tilde{\lambda}_{\rm u}P_{\rm ut}^{\frac{2}{\alpha}}\mathbb{E}[S_{\rm D}^{\frac{2}{\alpha}}]z^{\frac{2}{\alpha}}}{\alpha\sin\left(\frac{2\pi}{\alpha}\right)}\right)$$
$$\times\exp\left(-\tau_{\rm s}a_{\rm s}\,\mathcal{C}_{\alpha,K_{\rm s}}(zP_{\rm st},t)\left(\frac{zP_{\rm st}}{K_{\rm s}}\right)^{\delta}-\tau_{\rm m}a_{\rm m}\,\mathcal{C}_{\alpha,K_{\rm m}}(zP_{\rm st},t)\left(\frac{zP_{\rm mt}}{K_{\rm m}}\right)^{\delta}\right)f_{L_{\rm s}}(t)dtdz \tag{16}$$

$$R_{\rm s}^{\rm UL} = (1-\zeta_{\rm b})\int_0^\infty \frac{e^{-\sigma^2 z}}{z\ln 2}\left[1-\int_0^\infty\frac{f_{L_{\rm s}}(t)dt}{(1+zP_{\rm ut}/t)^{\Delta_{\rm s}}}\right]\exp\left\{-\tilde{\lambda}_{\rm u}\pi\mathbb{E}[S_{\rm D}^{\delta}]\int_0^\infty\frac{1-e^{-G_{\rm s}z}}{1+z^{-1}u^{\frac{1}{\delta}}/P_{\rm ut}}du\right.$$
$$\left.-\frac{\Gamma(1+\delta)\,\delta\pi^2 z^{\delta}}{\sin(\delta\pi)}\left[\frac{\tau_{\rm s}a_{\rm s}P_{\rm st}^{\delta}\prod_{i=1}^{K_{\rm s}-1}(i+\delta)}{\Gamma(K_{\rm s})K_{\rm s}^{\delta}}+\frac{\tau_{\rm m}a_{\rm m}P_{\rm mt}^{\delta}\prod_{i=1}^{K_{\rm m}-1}(i+\delta)}{\Gamma(K_{\rm m})K_{\rm m}^{\delta}}\right]\right\}dz. \tag{18}$$

energy efficiency gain over a one-tier network if the backhaul bandwidth is optimally allocated. Unless otherwise stated, the analytical expressions provided in this section are tight approximations of the actual data rates. For a better readability, most proofs and mathematical derivations have been relegated to the Appendix.

### *A. Analysis*

Under dynamic TDD [42], [43], transmissions are corrupted by DL interference from other MBSs and SAPs, and by UL interference from UEs that associated with other MBSs and SAPs. Specifically, the UL interference from UEs that associated with MBSs follow a homogeneous PPP with density $(1-\tau_{\rm m})\lambda_{\rm m}K_{\rm m}$, and similarly, the UL interference from UEs that associated with SAPs follow a homogeneous PPP with density $(1-\tau_{\rm s})\lambda_{\rm s}K_{\rm s}$. By the composition theorem [60], we have the UL interfering UEs follow a PPP with density $\tilde{\lambda}_{\rm u}=(1-\tau_{\rm m})\lambda_{\rm m}K_{\rm m}+(1-\tau_{\rm s})\lambda_{\rm s}K_{\rm s}$.

By noting that in practice, MBS can equip a large number of antennas, we use random matrix theory tools to obtain the DL rate on a macro cell link.

**Lemma 2:** *The downlink rate on a macro cell is given by* (11)*, where* $\delta=2/\alpha$, $a_{\rm m}=\lambda_{\rm m}\pi\mathbb{E}[S_{\rm D}^{\delta}]$, $a_{\rm s}=\lambda_{\rm s}\pi\mathbb{E}[S_{\rm D}^{\delta}]$, $\tilde{\lambda}_{\rm u}=(1-\tau_{\rm m})\lambda_{\rm m}K_{\rm m}+(1-\tau_{\rm s})\lambda_{\rm s}K_{\rm s}$*, while* $\nu_{\rm m}^{\rm D}$, $f_{L_{\rm m}}(t)$*, and* $\mathcal{C}_{\alpha,K}(z,t)$ *are given respectively as follows*

$$\nu_{\rm m}^{\rm D}=\frac{P_{\rm mt}(1-\beta_{\rm m})(G_{\rm m})^{\frac{\alpha}{2}}}{\beta_{\rm m}\Gamma\left(1+\frac{\alpha}{2}\right)}, \tag{12}$$

$$f_{L_{\rm m}}(t)=G_{\rm m}\delta x^{\delta-1}\exp\left(-G_{\rm m}x^{\delta}\right),\quad x\geq 0 \tag{13}$$

$$\mathcal{C}_{\alpha,K}(z,t)=\frac{2}{\alpha}\sum_{n=1}^{K}\binom{K}{n}\left[B\left(1;K-n+\frac{2}{\alpha},n-\frac{2}{\alpha}\right)\right.$$
$$\left.-B\left(\left(1+\frac{s}{tK}\right)^{-1};K-n+\frac{2}{\alpha},n-\frac{2}{\alpha}\right)\right] \tag{14}$$

*with* $G_{\rm m}=a_{\rm m}+a_{\rm s}\left(P_{\rm st}/P_{\rm mt}\right)^{\delta}$*, and* $B(x;y,z)=\int_0^x t^{y-1}(1-t)^{z-1}dt$ *the incomplete Beta function.*

*Proof:* See Appendix A. □

We note that in the downlink, due to the maximum received power association, interfering base station cannot be located closer to the typical user than the tagged base station, i.e., an exclusion region exists where the distance between a UE and the interfering base stations is bounded away from zero. However in the uplink, because of the PPP deployment assumption, an interfering base station can be located arbitrarily close to a typical MBS, i.e., the distance between a MBS and the interfering base stations can be arbitrarily small. In the following, we treat the latter as a composition of three independent PPPs with different spatial densities. We then obtain the macro cell uplink rate as follows.

**Lemma 3:** *The uplink rate on a macro cell is given by* (15)*, with* $\nu_{\rm m}^{\rm U}=(1-\beta_{\rm m})M_{\rm m}P_{\rm mt}$.

*Proof:* See Appendix B. □

Unlike the macro cell, due to the relatively small number of antennas at the SAPs, random matrix theory tools cannot be employed to calculate the rate on a small cell. We therefore use the effective channel distribution as follows.

**Lemma 4:** *The downlink rate on a small cell is given by* (16)*, where* $\Delta_{\rm s}=M_{\rm s}-K_{\rm s}+1$*, and* $f_{L_{\rm s}}(t)$ *is given as*

$$f_{L_{\rm s}}(t)=G_{\rm s}\delta t^{\delta-1}\exp\left(-G_{\rm s}t^{\delta}\right),\ t\geq 0 \tag{17}$$

$$R_{\rm b}^{\rm DL} =\frac{\zeta_{\rm b} M_{\rm s}}{K_{\rm s}}\int_0^\infty \frac{e^{-\sigma^2 z}}{z\ln 2}\left(1-e^{-z\nu_{\rm b}^{\rm D}}\right)\exp\left(-\frac{\Gamma\left(1+\delta\right)\delta\pi^2 z^\delta P_{\rm sb}^\delta}{\sin(\delta\pi)\Gamma(M_{\rm s})M_{\rm s}^\delta}\mathbb{E}[S_{\rm B}^\delta](1-\tau_{\rm b})\lambda_{\rm s}\prod_{i=1}^{M_{\rm s}-1}(i+\delta)\right)$$
$$\times\int_0^\infty \exp\left(-\tau_{\rm b} a_{\rm b}\mathcal{C}_{\alpha,K_{\rm b}M_{\rm s}}\left(zP_{\rm mb},t\right)\left(\frac{zP_{\rm mb}}{K_{\rm b}M_{\rm s}}\right)^\delta\right)f_{L_{\rm b}}(t)dtdz \tag{19}$$

$$R_{\rm b}^{\rm UL} = \frac{\zeta_{\rm b} M_{\rm s}}{K_{\rm s}}\int_0^\infty\int_0^\infty \frac{e^{-\sigma^2 z}}{z\ln 2}\left(1-e^{-z\nu_{\rm b}^{\rm U}/t}\right)\exp\left\{-\frac{\tau_{\rm b} a_{\rm b}\Gamma(1+\delta)\delta\pi^2 P_{\rm mb}^\delta z^\delta\prod_{i=1}^{K_{\rm b}M_{\rm s}-1}(i+\delta)}{\sin(\delta\pi)\left(M_{\rm s}K_{\rm b}\right)^\delta\Gamma\left(M_{\rm s}K_{\rm b}\right)}\right\}$$
$$\times\exp\left\{-\left(1-\tau_{\rm b}\right)a_{\rm b}K_{\rm b}\sum_{n=1}^{M_{\rm s}}\binom{M_{\rm s}}{n}\int_0^\infty\frac{\left(zu^{-1/\delta}P_{\rm sb}/M_{\rm s}\right)^n\left(1-e^{-a_{\rm b}u}\right)}{\left(1+zu^{-1/\delta}P_{\rm sb}/M_{\rm s}\right)^{M_{\rm s}}}du\right\}f_{L_{\rm b}}(t)dtdz \tag{22}$$

$$R_{\rm m,FDD}^{\rm DL} = \xi_{\rm D}(1-\zeta_{\rm b})\int_0^\infty\int_0^\infty\frac{e^{-\sigma^2 z}}{z\ln 2}\left(1-e^{-z\nu_{\rm m}^{\rm D}}\right)$$
$$\times\exp\left(-\tau_{\rm m} a_{\rm m}\mathcal{C}_{\alpha,K_{\rm m}}(zP_{\rm mt},t)\left(\frac{zP_{\rm mt}}{K_{\rm m}}\right)^\delta-\tau_{\rm s} a_{\rm s}\mathcal{C}_{\alpha,K_{\rm s}}(zP_{\rm mt},t)\left(\frac{zP_{\rm st}}{K_{\rm s}}\right)^\delta\right)f_{L_{\rm m}}(t)dtdz. \tag{25}$$

$$R_{\rm m,FDD}^{\rm UL} = (1-\zeta_{\rm b})(1-\xi_{\rm D})\int_0^\infty\int_0^\infty\frac{e^{-\sigma^2 z}}{z\ln 2}\left(1-e^{-z\nu_{\rm m}^{\rm U}/t}\right)$$
$$\times\exp\left(-\tilde{\lambda}_{\rm u}\pi\mathbb{E}[S_{\rm D}^\delta]\int_0^\infty\frac{1-e^{-G_{\rm m}u}}{1+z^{-1}u^{\frac{1}{\delta}}/P_{\rm ut}}du\right)f_{L_{\rm U}}(t)dtdz \tag{26}$$

---

*with* $G_{\rm s}=a_{\rm s}+a_{\rm m}\left(P_{\rm mt}/P_{\rm st}\right)^\delta$.

*Proof:* See Appendix C. □

Following a similar approach as the one in Lemma 3, we can obtain the uplink rate on a small cell.

**Lemma 5:** *The uplink rate on a small cell is given by* (18) *on the top of previous page.*

*Proof:* The proof is similar to the one in Lemma 3 and it is omitted. □

We now derive downlink and uplink rates on the wireless backhaul of a heterogeneous network as follows.

**Lemma 6:** *The downlink rate on the wireless backhaul is given by* (19)*, where* $a_{\rm b}=\lambda_{\rm m}\pi\mathbb{E}[S_{\rm B}^\delta]$*,* $f_{L_{\rm b}}(t)$ *and* $\nu_{\rm b}^{\rm D}$ *are given as*

$$f_{L_{\rm b}}(t) = a_{\rm b}\delta t^{\delta-1}\exp(-a_{\rm b}t^\delta),\quad t>0 \tag{20}$$
$$\nu_{\rm b}^{\rm D} = \frac{P_{\rm mb}(1-\beta_{\rm b})a_{\rm b}^\delta}{\beta_{\rm b}\Gamma(1+1/\delta)}. \tag{21}$$

*Proof:* See Appendix D. □

**Lemma 7:** *The uplink rate on the wireless backhaul is given by* (22)*, where* $\nu_{\rm b}^{\rm U}=(1-\beta_{\rm b})M_{\rm m}P_{\rm sb}$.

*Proof:* See Appendix E. □

**Remark 1:** *In Lemma 6 and Lemma 7, we use ZF as the precoding and receiving scheme, which is suboptimal compared to the block diagonalization (BD) and can result in a lower data rate achievable on the wireless backhaul [56], [57]. However, the rate achievable under ZF is tractable, whereas to the best of the authors' knowledge no closed form expression is available for the rate achievable by BD. Furthermore, we show in Fig. 2 that the rate gap is limited, and that the rates under BD and ZF follow a similar trend. Therefore, our findings on the energy efficiency tradeoffs remain valid irrespective of the scheme used.*

By combining the previous results, we can now write the data rate per area in a heterogeneous network with wireless backhaul.

**Lemma 8:** *The sum rate per area in a heterogeneous network with wireless backhaul is given by*

$$\mathcal{R} = B\Big(K_{\rm m}\lambda_{\rm m}+K_{\rm s}\lambda_{\rm s}\Big)\Big\{A_{\rm m}\Big[\tau_{\rm m}R_{\rm m}^{\rm DL}+(1-\tau_{\rm m})R_{\rm m}^{\rm UL}\Big]$$
$$+A_{\rm s}\Big[\tau_{\rm s}\min\Big\{R_{\rm s}^{\rm DL},R_{\rm b}^{\rm DL}\Big\}+(1-\tau_{\rm s})\min\Big\{R_{\rm s}^{\rm UL},R_{\rm b}^{\rm UL}\Big\}\Big]\Big\} \tag{23}$$

*where* $B$ *is the total available bandwidth, and* $R_{\rm m}^{\rm DL}$*,* $R_{\rm m}^{\rm UL}$*,* $R_{\rm s}^{\rm DL}$*,* $R_{\rm s}^{\rm UL}$*,* $R_{\rm b}^{\rm DL}$*, and* $R_{\rm b}^{\rm UL}$ *are given in* (11), (15), (16), (18), (19), *and* (22), *respectively.*

*Proof:* See Appendix F. □

We finally obtain the energy efficiency of a heterogeneous network with wireless backhaul, defined as the number of bits successfully transmitted per joule of energy spent.

**Theorem 1:** *The energy efficiency* $\eta$ *of a heterogeneous network with wireless backhaul is given by*

$$\eta = \frac{B\left(K_{\rm m}\lambda_{\rm m}+K_{\rm s}\lambda_{\rm s}\right)}{P_{\rm m}\lambda_{\rm m}+P_{\rm s}\lambda_{\rm s}+P_{\rm b}\lambda_{\rm m}}\Big(A_{\rm m}\Big[\tau_{\rm m}R_{\rm m}^{\rm DL}+(1-\tau_{\rm m})R_{\rm m}^{\rm UL}\Big]$$
$$+A_{\rm s}\Big[\tau_{\rm s}\min\Big\{R_{\rm s}^{\rm DL},R_{\rm b}^{\rm DL}\Big\}+(1-\tau_{\rm s})\min\Big\{R_{\rm s}^{\rm UL},R_{\rm b}^{\rm UL}\Big\}\Big]\Big). \tag{24}$$

*Proof:* The result follows from Lemma 1 and Lemma 8 and by noting that the energy efficiency is obtained as the ratio between the data rate per area and the power consumption per area. □

$$R_{\mathrm{s,FDD}}^{\mathrm{DL}}=\xi_{\mathrm{D}}(1-\zeta_{\mathrm{b}})\int_0^{\infty}\!\!\int_0^{\infty}\frac{e^{-\sigma^2 z}}{z\ln 2}\left(1-\frac{1}{(1+zP_{\mathrm{st}}t^{-1}/K_{\mathrm{s}})^{\Delta_{\mathrm{s}}}}\right)$$

$$\times\exp\left(-\tau_{\mathrm{s}}a_{\mathrm{s}}\,\mathcal{C}_{\alpha,K_{\mathrm{s}}}(zP_{\mathrm{st}},t)\left(\frac{zP_{\mathrm{st}}}{K_{\mathrm{s}}}\right)^{\delta}-\tau_{\mathrm{m}}a_{\mathrm{m}}\,\mathcal{C}_{\alpha,K_{\mathrm{m}}}(zP_{\mathrm{st}},t)\left(\frac{zP_{\mathrm{mt}}}{K_{\mathrm{m}}}\right)^{\delta}\right)f_{L_{\mathrm{s}}}(t)dtdz. \tag{28}$$

$$R_{\mathrm{s,FDD}}^{\mathrm{UL}}=(1-\zeta_{\mathrm{b}})(1-\xi_{\mathrm{D}})\int_0^{\infty}\frac{e^{-\sigma^2 z}}{z\ln 2}\left[1-\int_0^{\infty}\frac{f_{L_{\mathrm{U}}}(t)dt}{(1+zP_{\mathrm{ut}}/t)^{\Delta_{\mathrm{s}}}}\right]\exp\left(-\tilde{\lambda}_{\mathrm{u}}\pi\mathbb{E}[S_{\mathrm{D}}^{\delta}]\int_0^{\infty}\frac{1-e^{-G_{\mathrm{s}}u}}{1+z^{-1}u^{\frac{1}{\delta}}/P_{\mathrm{ut}}}du\right)dz \tag{29}$$

$$R_{\mathrm{b,FDD}}^{\mathrm{DL}}=\frac{\xi_{\mathrm{B}}\zeta_{\mathrm{b}}M_{\mathrm{s}}}{K_{\mathrm{s}}\ln 2}\int_0^{\infty}\!\!\int_0^{\infty}\frac{\left(1-e^{-z\nu_{\mathrm{b}}^{\mathrm{D}}}\right)}{ze^{\sigma^2 z}}\exp\left(-\tau_{\mathrm{b}}a_{\mathrm{b}}\mathcal{C}_{\alpha,K_{\mathrm{b}}M_{\mathrm{s}}}(zP_{\mathrm{mb}},t)\left(\frac{zP_{\mathrm{mb}}}{K_{\mathrm{b}}M_{\mathrm{s}}}\right)^{\delta}\right)f_{L_{\mathrm{b}}}(t)dtdz \tag{30}$$

$$R_{\mathrm{b,FDD}}^{\mathrm{UL}}=\frac{(1-\xi_{\mathrm{B}})\,\zeta_{\mathrm{b}}M_{\mathrm{s}}}{K_{\mathrm{s}}}\int_0^{\infty}\frac{e^{-\sigma^2 z}}{z\ln 2}\int_0^{\infty}\left(1-e^{-z\nu_{\mathrm{b}}^{\mathrm{U}}/t}\right)f_{L_{\mathrm{U}}}(t)dt$$

$$\times\exp\left\{-(1-\tau_{\mathrm{b}})\,a_{\mathrm{b}}K_{\mathrm{b}}\sum_{n=1}^{M_{\mathrm{s}}}\binom{M_{\mathrm{s}}}{n}\int_0^{\infty}\frac{\left(zu^{-1/\delta}P_{\mathrm{sb}}/M_{\mathrm{s}}\right)^{n}(1-e^{-a_{\mathrm{b}}u})}{\left(1+zu^{-1/\delta}P_{\mathrm{sb}}/M_{\mathrm{s}}\right)^{M_{\mathrm{s}}}}du\right\}dz \tag{31}$$

---

Equation (24) quantifies how all the key features of a heterogeneous network, i.e., interference, deployment strategy, and capability of the wireless infrastructure components, affect the energy efficiency when a wireless backhaul is used to forward traffic into the core network. Several numerical results based on (24) will be shown in Section V to give more practical insights into the energy-efficient design of a heterogeneous network with wireless backhaul. In Section IV-C, we provide simulations to validate the analysis presented in this section.

### B. Rate Analysis of FDD network

Our results are general and hold under both time division duplex (TDD) and frequency division duplex (FDD). In fact, TDD and FDD are equivalent in that they all divide up the spectrum orthogonally [61]. In this section, we show the extension of our framework to the case of FDD, where a portion $\xi_{\mathrm{D}}$ of the radio access spectrum is assigned to the downlink, and the remaining fraction $(1-\xi_{\mathrm{D}})$ is assigned to the uplink. Similarly, on the wireless backhaul, a fraction $\xi_{\mathrm{B}}$ is reserved for the downlink, and the remaining fraction $(1-\xi_{\mathrm{B}})$ is reserved for the uplink. In addition, in order to increase the spectral efficiency, we consider a decoupled UL/DL association, where the downlink UEs are associated with the base station that provides the largest received power, and the uplink UEs are associated with the closest MBSs or SAPs. As such, the rate of DL macro cell, UL macro cell, DL small cell, UL small cell, and the wireless backhaul can be written as follows.

**Lemma 9:** *Under FDD, the downlink rate on a macro cell is given by* (25) *in the previous page.*

*Proof:* See Appendix G. □

**Lemma 10:** *Under FDD, the uplink rate on a macro cell is given by* (26) *in the previous page, where* $f_{L_{\mathrm{U}}}(t)$ *is given as*

$$f_{L_{\mathrm{U}}}(t)=\delta G t^{\delta-1}\exp\left(-Gt^{\delta}\right),\quad G=(\lambda_{\mathrm{s}}+\lambda_{\mathrm{m}})\,\pi\mathbb{E}[S_{\mathrm{D}}^{\delta}]. \tag{27}$$

*Proof:* The proof is similar to the one in Lemma 3 and it is omitted. □

**Lemma 11:** *Under FDD, the downlink rate on a small cell is given by* (28)*.*

*Proof:* The proof is similar to the one in Lemma 4 and it is omitted. □

**Lemma 12:** *Under FDD, the uplink rate on a small cell is given by* (29)*.*

*Proof:* The proof is similar to the one in Lemma 5 and it is omitted. □

**Lemma 13:** *Under FDD, the downlink rate on the wireless backhaul is given by* (30)*.*

*Proof:* The proof is similar to the one in Lemma 6 and it is omitted. □

**Lemma 14:** *Under FDD, the uplink rate on the wireless backhaul is given by* (31)*.*

*Proof:* The proof is similar to the one in Lemma 7 and it is omitted. □

### C. Validation

We now show simulation results that confirm the accuracy of the analysis provided in this section. In our simulations, all cells operate under dynamic TDD, the locations of MBSs, SAPs, and UEs are generated as PPPs, and the typical UE is located at the origin. We use the following values for the number of antennas and the transmit power: $M_{\mathrm{m}}=100$, $M_{\mathrm{s}}=4$, $P_{\mathrm{mt}}=47.8$dBm, and $P_{\mathrm{st}}=23.7$dBm.

In Fig. 2 we compare the rate on the wireless backhaul under block diagonalization (BD) and zero forcing (ZF), respectively, with different numbers of SAPs. Although ZF achieves a lower rate than BD, the rate gap is limited as the antenna number grows, and the rates under BD and ZF follow a similar trend. Therefore, the conclusions drawn in this paper on the energy efficiency tradeoffs remain valid irrespective of the scheme used.

Fig. 3 compares the simulated macro cell downlink rate to the analytical result obtained in Lemma 2 with different antenna numbers at the MBS. The downlink rate is plotted versus the transmit power at the MBSs. The figure shows that analytical results and simulations fairly well match, thus confirming the accuracy of Lemma 2.

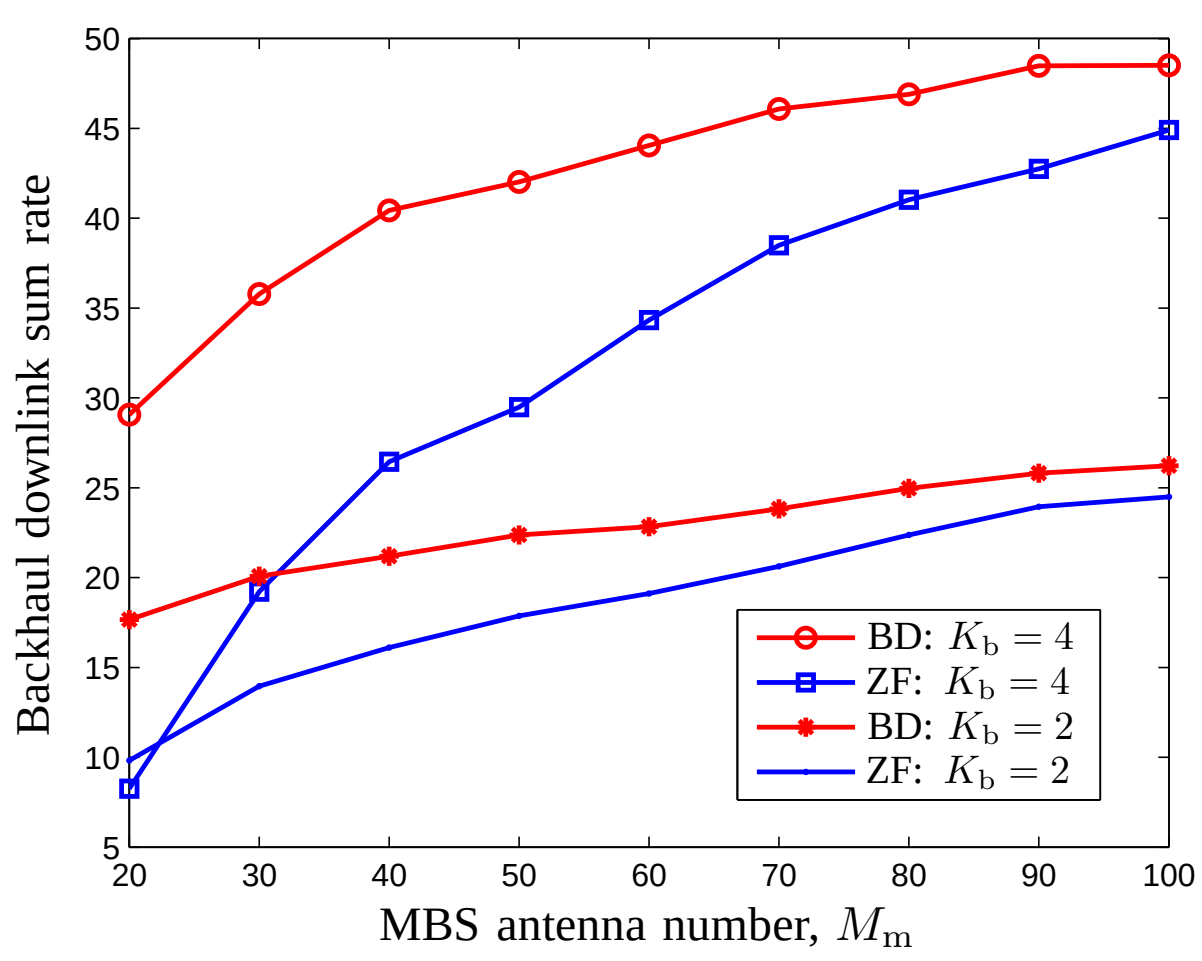

Fig. 2. Downlink rate on the wireless backhaul, by using block diagonalization and zero forcing, with different numbers of SAPs.

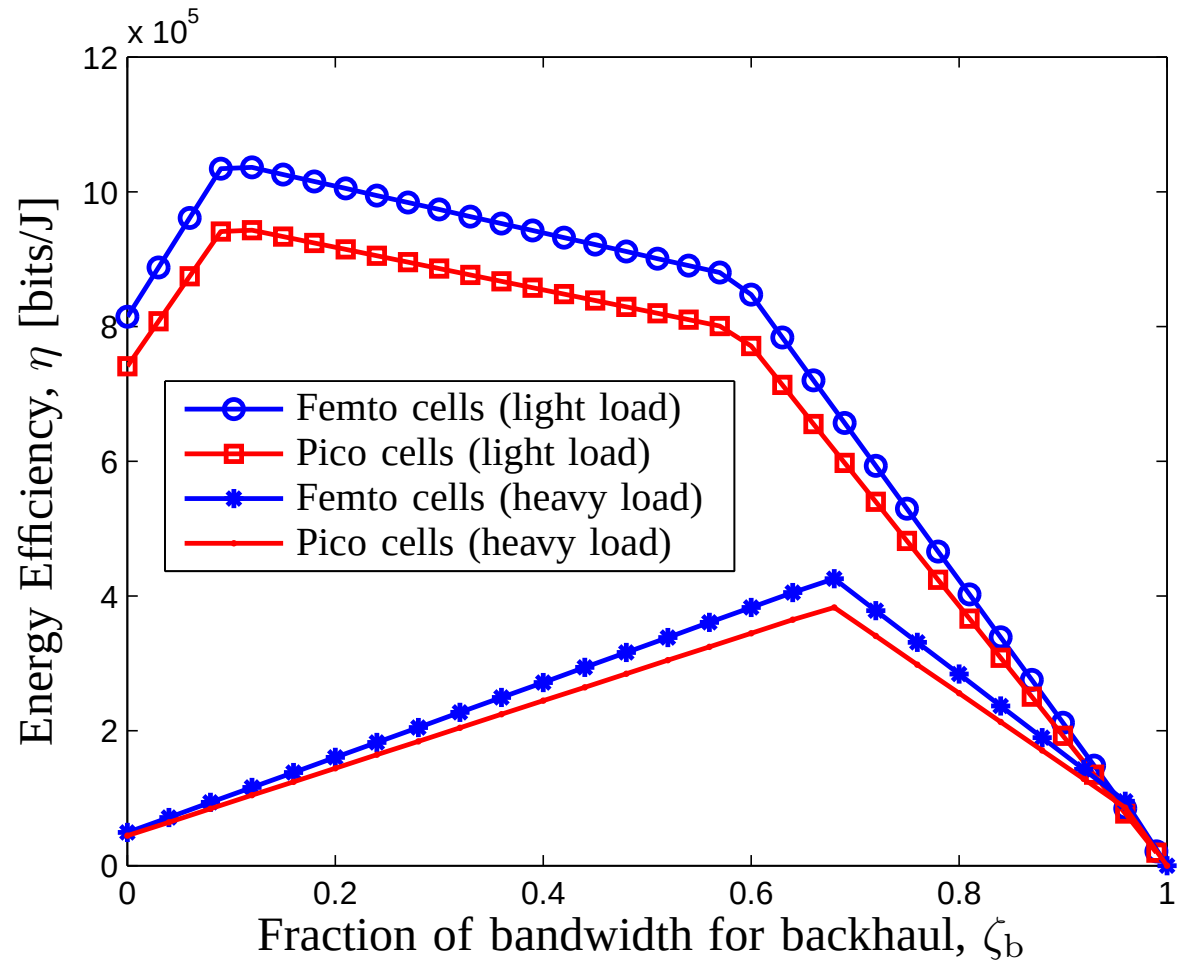

Fig. 4. Energy efficiency of a heterogeneous network that uses pico cells and femto cells, respectively, versus fraction of bandwidth $\zeta_{\rm b}$ allocated to the backhaul, under different load conditions.

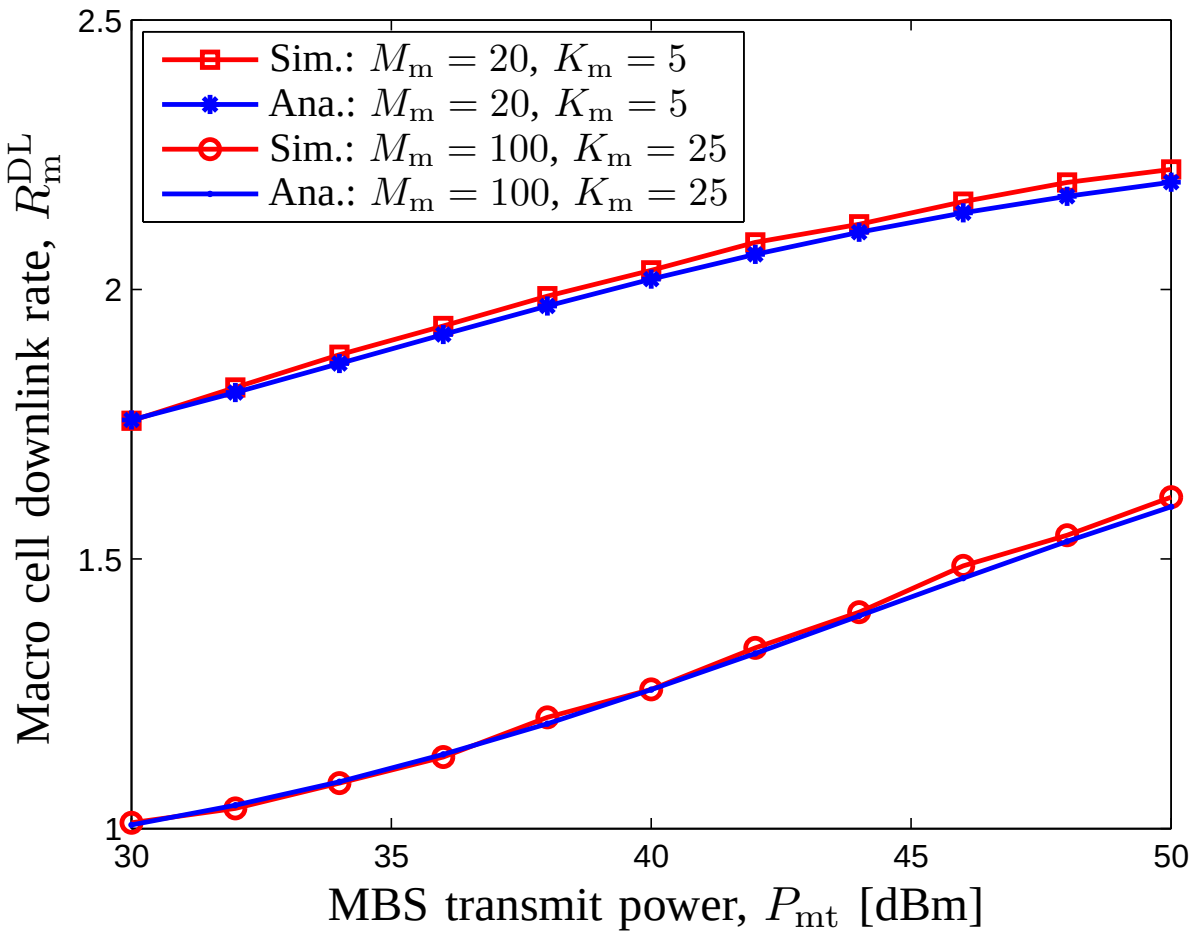

Fig. 3. Comparison of the simulations and numerical results for macro cell downlink rate.

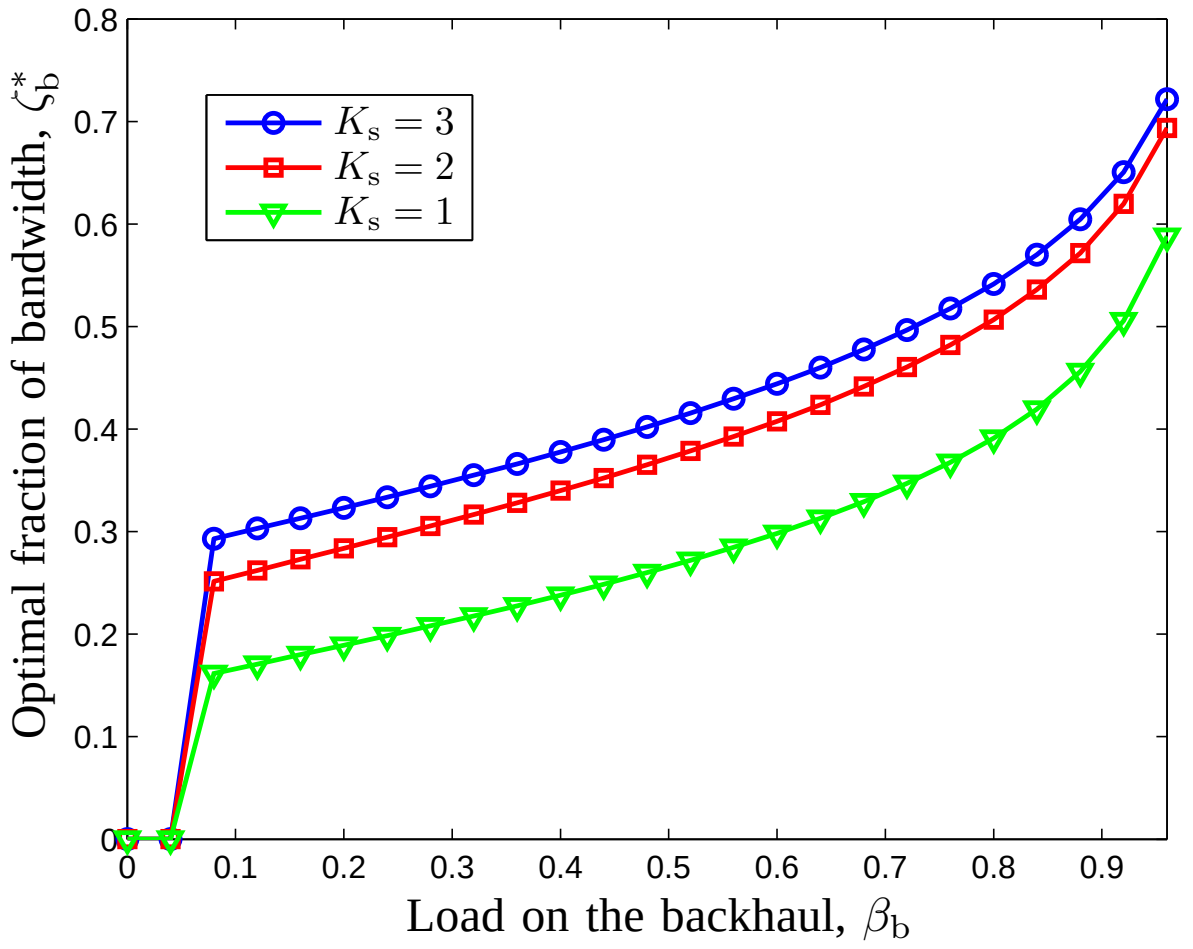

Fig. 5. Optimal fraction of bandwidth to be allocated to the backhaul versus load on the backhaul, for various values of the number of UEs per SAP, $K_{\rm s}$.

## V. Numerical Results

In this section, we provide numerical results to show how the energy efficiency is affected by various network parameters and to give insights into the optimal design of a heterogeneous network with wireless backhaul. As an example, we consider two different deployment scenarios, namely (i) a dense deployment of low-power SAPs with a small number of antennas, here denoted as *femto cells*, and (ii) a less dense deployment of larger and more powerful SAPs, here denoted as *pico cells*, and we refer to *light load* and *heavy load* conditions as the ones of a network with $\beta_{\rm m}=\beta_{\rm s}=\beta_{\rm b}=0.25$ and $0.9\leq\beta_{\rm m}$, $\beta_{\rm s}$, $\beta_{\rm b}<1$, respectively. We consider a network operating at 2GHz, we set the path loss exponent to $\alpha=3.8$ to model an urban scenario, the shadowing $S_{\rm B}$ and $S_{\rm D}$ are set to be lognormal distributed as $S_{\rm B}=10^{\frac{X_{\rm B}}{10}}$ and $S_{\rm D}=10^{\frac{X_{\rm D}}{10}}$, where $X_{\rm B}\sim N(0,\sigma_{\rm B}^2)$ and $X_{\rm D}\sim N(0,\sigma_{\rm D}^2)$, with $\sigma_{\rm B}=3$dB and $\sigma_{\rm D}=6$dB, respectively [62]. In addition, we set the backhaul transmit power equal to the radio access power, i.e., $P_{\rm mb}=P_{\rm mt}, P_{\rm sb}=P_{\rm st}$. All other system and power consumption parameters are set as follows: $P_{\rm mt}=47.8$dBm, for pico cell SAPs $P_{\rm st}=30$dBm, for femto cell SAP $P_{\rm st}=23.7$dBm, $P_{\rm ut}=17$dBm, $P_{\rm ma}=1$W, for pico cell SAPs $P_{\rm sa}=0.8$W, for femto cell SAP $P_{\rm sa}=0.8$W, $P_{\rm ua}=0.1$W [32]; $P_{\rm mf}=225$W, for pico cell SAPs $P_{\rm sf}=7.3$W, for femto cell SAPs $P_{\rm sf}=5.2$W [58], [59]; $P_{\rm me}=0.1$W/Gb, $P_{\rm md}=0.8$W/Gb, $P_{\rm se}=0.2$W/Gb, $P_{\rm sd}=1.6$W/Gb, $P_{\rm ue}=0.3$W/Gb, $P_{\rm md}=2.4$W/Gb [49].

In Fig. 4, we compare the energy efficiency of heterogeneous networks that use pico cells and femto cells, respectively, under various load conditions and for different portions of the bandwidth allocated to the wireless backhaul. The figure shows that femto cell and pico cell deployments exhibit similar performance in terms of energy efficiency. Moreover, Fig. 4 shows that the energy efficiency of the network is highly sensitive to the portion of bandwidth allocated to the backhaul, and that there is an optimal value of $\zeta_{\mathrm{b}}$ which maximizes the energy efficiency of the HetNet. The optimal value of $\zeta_{\mathrm{b}}$ is not affected by the network infrastructure, i.e., it is the same for pico cells and femto cells. However, the optimal $\zeta_{\mathrm{b}}$ increases as the load on the network increases. In fact, when more UEs associate with each SAPs, more data need to be forwarded from MBSs to SAPs through the wireless backhaul in order to meet the rate demand. In summary, the figure shows that irrespective of the deployment strategy, an optimal backhaul bandwidth allocation that depends on the network load can be highly beneficial to the energy efficiency of a heterogeneous network.

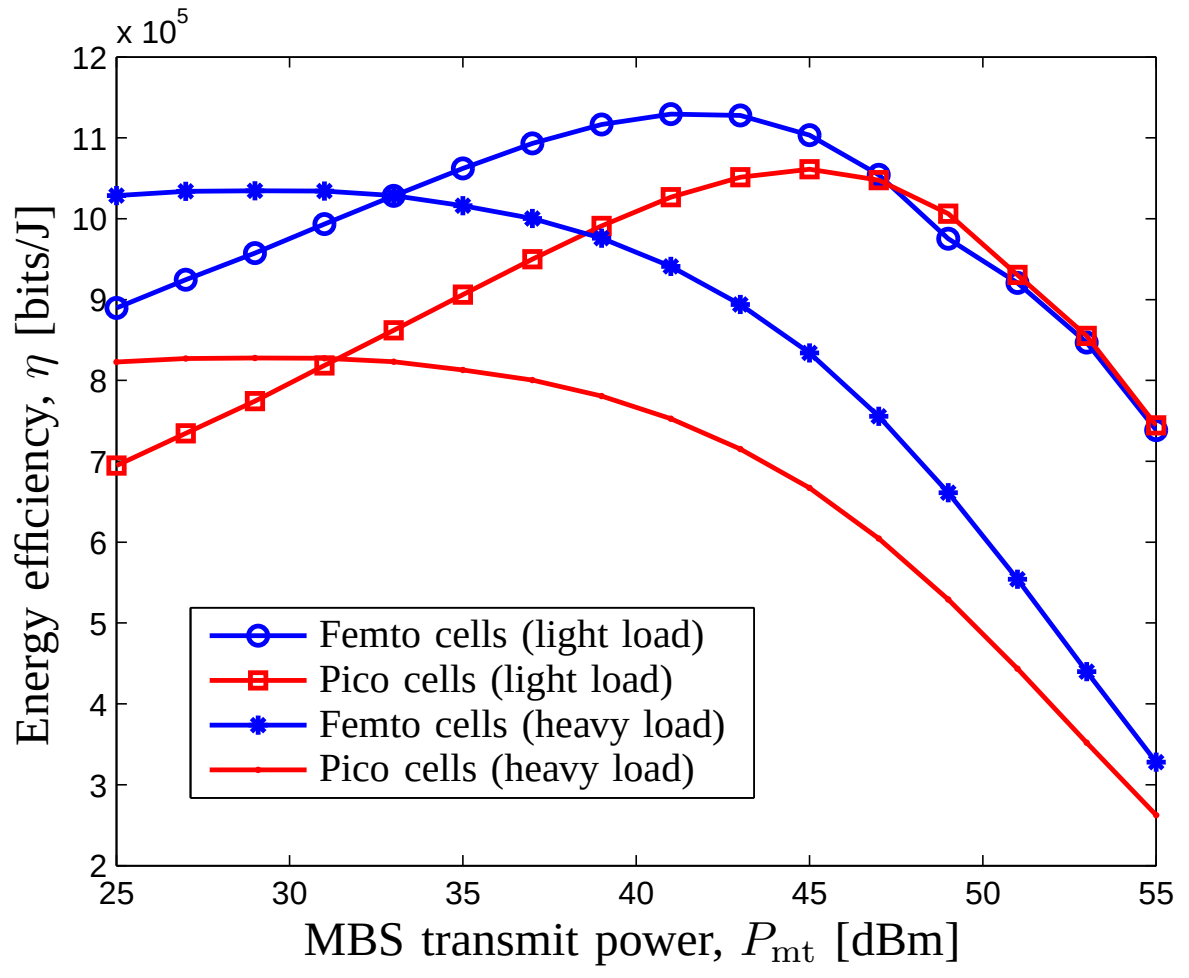

Fig. 6. Energy efficiency of a heterogeneous network that uses pico cells and femto cells, respectively, versus power allocated to the backhaul, under different load conditions.

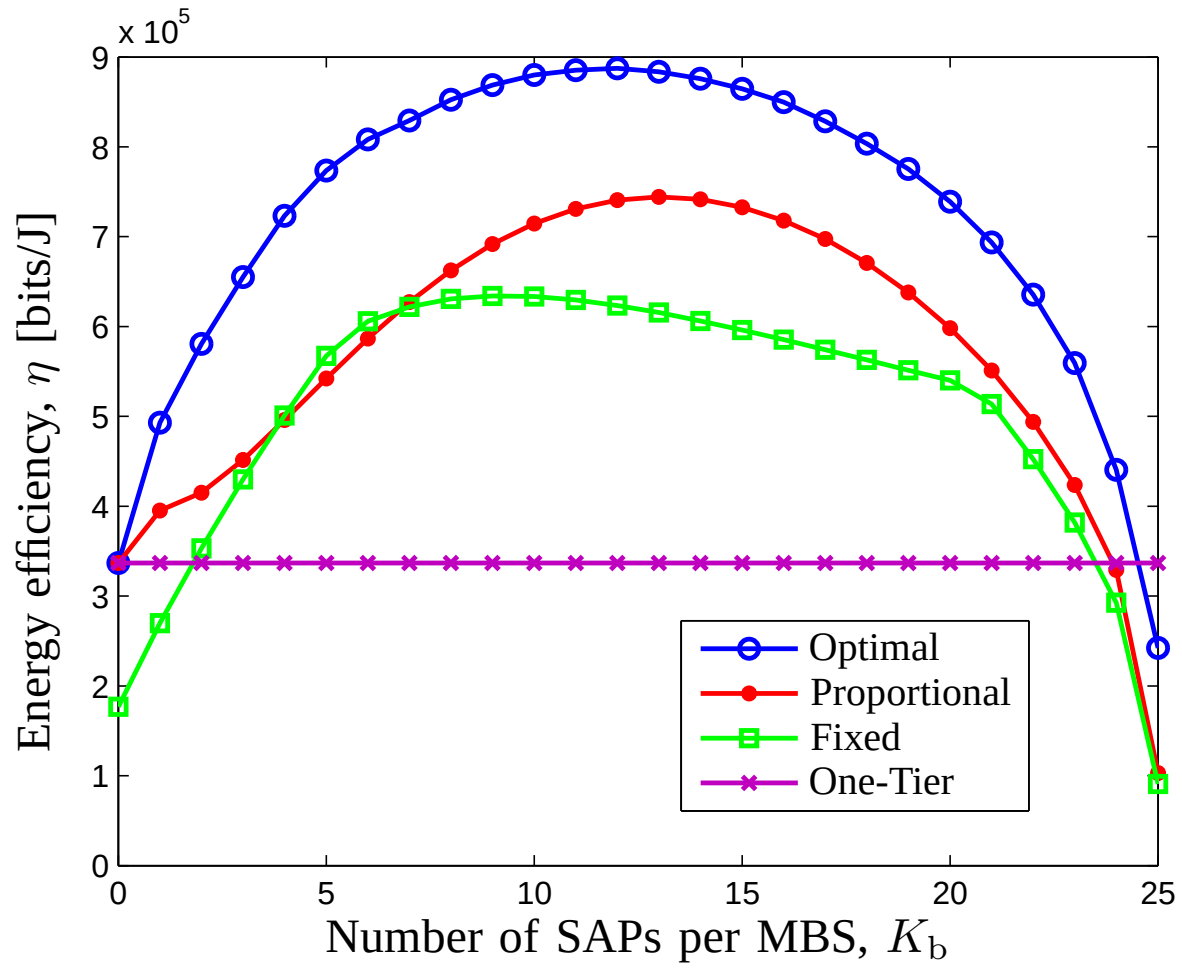

Fig. 7. Energy efficiency versus number of SAPs per MBS under various bandwidth allocation schemes.

In Fig. 5, we plot the optimal value $\zeta_{\mathrm{b}}^*$ for the fraction of bandwidth to be allocated to the backhaul as a function of the load on the backhaul $\beta_{\mathrm{b}}$. We consider femto cell deployment for three different values of the number of UEs per SAP, $K_{\mathrm{s}}$. Consistently with Fig. 4, this figure shows that the optimal fraction of bandwidth $\zeta_{\mathrm{b}}^*$ to be allocated to the wireless backhaul increases as $\beta_{\mathrm{b}}$ or $K_{\mathrm{s}}$ increase, since the load on the wireless backhaul becomes heavier and more resources are needed to meet the data rate demand.

In Fig. 6, we plot the energy efficiency of the HetNet as a function of the MBS transmit power under different deployment strategies and load conditions. The figure shows that the energy efficiency is sensitive to the MBS transmit power, and that there is an optimal value for the transmit power, given by a tradeoff between the data rate that the wireless backhaul can support and the power consumption incurred. Under spatial multiplexing, the data rate of the network is affected by the number of scheduled UEs per base station antenna, which we denote as the network load. As a consequence, the network load affects the data rate, and in turn affects the energy efficiency.

In Fig. 7, we plot the energy efficiency of the network versus the number of SAPs per MBS. We consider four scenarios: (i) optimal bandwidth allocation, where the fraction of bandwidth $\zeta_{\mathrm{b}}$ for the backhaul is chosen as the one that maximizes the overall energy efficiency; (ii) proportional bandwidth allocation, where the fraction of bandwidth allocated to the backhaul is equal to the fraction of load on the backhaul, i.e., $\zeta_{\mathrm{b}} = \frac{K_{\mathrm{b}} K_{\mathrm{s}}}{K_{\mathrm{m}} + K_{\mathrm{b}} K_{\mathrm{s}}}$ [41]; (iii) fixed bandwidth allocation, where the bandwidth is equally divided between macro-and-small-cell links and wireless backhaul, i.e., $\zeta_{\mathrm{b}} = 0.5$; and (iv) one-tier cellular network, where no SAPs or wireless backhaul are used at all, and all the bandwidth is allocated to the macro cell link, i.e., $\zeta_{\mathrm{b}} = 0$. Fig. 7 shows that in a two-tier heterogeneous network there is an optimal number of SAPs associated to each MBS via the wireless backhaul that maximizes the energy efficiency. Such number is given by a tradeoff between the data rate that the SAPs can provide to the UEs and the total power consumption. This figure also indicates that if the wireless backhaul is not supported well, a two-tier HetNet with wireless backhaul can be worse than a single tier cellular network in terms of energy efficiency. However, when the backhaul bandwidth is optimally allocated, the HetNet can achieve a significant energy efficiency gain over a one tier deployment.

## VI. Concluding Remarks

In this work, we undertook an analytical study for the energy-efficient design of heterogeneous networks with a wireless backhaul. We used a general model that accounts for uplink and downlink transmissions, spatial multiplexing, and resource allocation between radio access links and backhaul. Our results revealed that, irrespective of the deployment strategy, it is critical to control the network load in order to maintain a high energy efficiency. Moreover, a two-tier heterogeneous network with wireless backhaul can achieve a significant energy efficiency gain over a one-tier deployment, as long as the bandwidth division between radio access links and wireless backhaul is optimally designed.

The framework provided in this paper allows to explicitly characterize the power consumption of the HetNet due to the signal processing operations in macro cells, small cells, and wireless backhaul, as well as the data rates and ultimately the energy efficiency of the whole network. More generally, our work helps to understand how all the key features of a heterogeneous network, i.e., interference, load, deployment strategy, and capability of the wireless infrastructure components, affect the energy efficiency when a wireless backhaul is used to forward traffic into the core network.

This paper considered the current state-of-the-art co-channel deployment of small cells with the macro cell tier. In the near future, an orthogonal, ultra-dense deployment of small cells could be used to further boost the network capacity by targeting static users. Investigating up to what extent the wireless backhaul capability can support such ultra-dense topology, and designing idle-mode mechanisms for an energy-efficient and sustainable ultra-dense deployment are regarded as concrete directions for future work.

## APPENDIX

### A. Proof of Lemma 2

The channel matrix between a MBS to its $K_{\rm m}$ associated UEs can be written as $\hat{\mathbf{H}} = \mathbf{L}^{\frac{1}{2}}\mathbf{H}$, where $\mathbf{L} = {\rm diag}\{L_1^{-1}, ..., L_{K_{\rm m}}^{-1}\}$, with $L_i = r_i^{\alpha}/S_i$ being the path loss from the MBS to its $i$-th UE, where $r_i$ is the corresponding distance and $S_i$ denotes the shadowing, $\mathbf{H} = [\mathbf{h}_1, ..., \mathbf{h}_{K_{\rm m}}]^{\rm T}$ is the $K_{\rm m} \times M_{\rm m}$ small scale fading matrix, with $\mathbf{h}_i \sim \mathcal{CN}(\mathbf{0}, \mathbf{I})$. The ZF precoder is then given by $\mathbf{W} = \xi\hat{\mathbf{H}}^*(\hat{\mathbf{H}}\hat{\mathbf{H}}^*)^{-1}$, where $\xi^2 = 1/{\rm tr}[(\hat{\mathbf{H}}^*\hat{\mathbf{H}})^{-1}]$ normalizes the transmit power [62]. In the following, we use the notation $\Phi^{\rm U}$ as $\Phi^{\rm D}$ to denote the subsets of $\Phi$ that transmit in uplink and downlink, respectively, we further denote $\mathcal{U}_x$ as the set of UEs that are associated with access point $x$, and denote $\hat{x}$ as the transmitter that locates closest to the origin. Since the locations of MBSs and SAPs follow a stationary PPP, we can apply the Slivnyark's theorem [60], which implies that it is sufficent to evaluate the signal-to-interference-plus-noise ratio (SINR) of a typical UE at the origin. As such, by noticing that under dynamic TDD, every wireless link experiences interference from the downlink transmitting MBSs and SAPs, and from the uplink transmitting UEs, the downlink SINR between a typical UE at the origin and its serving MBS can be written as

$$\gamma_{\rm m}^{\rm DL} = \frac{P_{\rm mt}|\mathbf{h}_{\hat{x}_{\rm m},o}^*\mathbf{w}_{\hat{x}_{\rm m},o}|^2 L_{\hat{x}_{\rm m},o}^{-1}}{I_{\rm oc}^{\rm mu} + I_{\rm u} + \sigma^2} \tag{28}$$

where $\mathbf{h}_{\hat{x}_{\rm m},o}$ is the small scale fading, $\mathbf{w}_{\hat{x}_{\rm m},o}$ is the ZF precoding vector, $L_{\hat{x}_{\rm m},o}$ denotes the corresponding path loss, while $I_{\rm oc}^{\rm mu}$ is the aggregate interference from other cells to the MBS UE, and $I_{\rm u}$ denotes the interference from UEs, respectively given as follows

$$I_{\rm oc}^{\rm mu} = \sum_{x_{\rm m} \in \Phi_{\rm m}^{\rm D} \backslash \hat{x}_{\rm m}} \frac{P_{\rm mt} g_{x_{\rm m},o}}{K_{\rm m} L_{x_{\rm m},o}} + \sum_{x_{\rm s} \in \Phi_{\rm s}^{\rm D}} \frac{P_{\rm st} g_{x_{\rm s},o}}{K_{\rm s} L_{x_{\rm s},o}} \tag{29}$$

and

$$I_{\rm u} = \sum_{x_{\rm u} \in \Phi_{\rm u}^{\rm U}} \frac{P_{\rm ut}|h_{x_{\rm u},o}|^2}{L_{x_{\rm u},o}} \tag{30}$$

whereas $g_{x_{\rm m},o}$ and $g_{x_{\rm s},o}$ represent the effective small-scale fading from the interfering MBS $x_{\rm m}$ and SAP $x_{\rm s}$ to the origin, respectively, given by [63]

$$g_{x_{\rm m},o} = \sum_{u \in \mathcal{U}_{x_{\rm m}}} K_{\rm m}|\mathbf{h}_{x_{\rm m},o}^*\mathbf{w}_{x_{\rm m},u}|^2 \sim \Gamma(K_{\rm m}, 1) \tag{31}$$

and

$$g_{x_{\rm s},o} = \sum_{u \in \mathcal{U}_{x_{\rm s}}} K_{\rm s}|\mathbf{h}_{x_{\rm s},o}^*\mathbf{w}_{x_{\rm s},u}|^2 \sim \Gamma(K_{\rm s}, 1). \tag{32}$$

By conditioning on the interference, when $K_{\rm m}, M_{\rm m} \to \infty$ with $\beta_{\rm m} = K_{\rm m}/M_{\rm m} < 1$, the SINR under ZF precoding converges to [45]

$$\gamma_{\rm m}^{\rm DL} \to \bar{\gamma}_{\rm m}^{\rm DL} = \frac{P_{\rm mt} M_{\rm m}}{(I_{\rm oc}^{\rm mu} + I_{\rm u} + \sigma^2)\sum_{j=1}^{K_{\rm m}} e_j^{-1}}, \quad a.s. \tag{33}$$

where $e_i$ is the solution of the fixed point equation

$$\frac{L_{\hat{x}_{\rm m},u_i}^{-1}}{e_i} = 1 + \frac{J}{M_{\rm m}}, \quad i = 1, 2, ..., K_{\rm m} \tag{34}$$

with $J = \sum_{j=1}^{K_{\rm m}} L_{\hat{x}_{\rm m},u_j}^{-1} e_j^{-1}$. By summing (34) over $i$ we obtain

$$J = K_{\rm m} + \frac{K_{\rm m}}{M_{\rm m}} J. \tag{35}$$

Solving the equation above results in $J = K_{\rm m}M_{\rm m}/(M_{\rm m} - K_{\rm m})$, and by substituting the value of $J$ into (34) we can have

$$\frac{1}{\bar{e}_i} = \frac{M_{\rm m}}{M_{\rm m} - K_{\rm m}} \cdot L_{\hat{x}_{\rm m},u_i} \tag{36}$$

which substituted into (33) yields

$$\bar{\gamma}_{\rm m}^{\rm DL} = \frac{(1 - \beta_{\rm m}) M_{\rm m} P_{\rm mt}}{(I_{\rm oc}^{\rm mu} + I_{\rm u} + \sigma^2)\sum_{j=1}^{K_{\rm m}} L_{\hat{x}_{\rm m},u_j}}. \tag{37}$$

Notice that $\{L_{\hat{x}_{\rm m},u_j}\}_{j=1}^{K_{\rm m}}$ is an independent independent and identically distributed (i.i.d.) sequence with finite first moment, given by

$$\mathbb{E}\left[L_{\hat{x}_{\rm m},u_j}\right] = \Gamma\left(1 + \frac{1}{\delta}\right) G_{\rm m}^{-1} < \infty,$$

by applying the strong law of large numbers (SLLN) to (37), we have

$$\bar{\gamma}_{\rm m}^{\rm DL} \to \frac{(1 - \beta_{\rm m}) G_{\rm m}^{1/\delta} P_{\rm mt}}{\beta_{\rm m}\Gamma\left(1 + \frac{1}{\delta}\right)(I_{\rm oc}^{\rm mu} + I_{\rm u} + \sigma^2)}, \quad a.s. \tag{38}$$

As such, using the continuous mapping theorem and the lemma in [64], we can compute the ergodic rate as

$$\begin{aligned} \mathbb{E}\left[\log_2\left(1 + \bar{\gamma}_{\rm m}^{\rm DL}\right)\right] &= \frac{1}{\ln 2}\mathbb{E}\left[\ln\left(1 + \frac{\nu_{\rm m}^{\rm D}}{I_{\rm oc}^{\rm mu} + I_{\rm u} + \sigma^2}\right)\right] \\ &= \int_0^\infty \frac{e^{-\sigma^2 z}}{z \ln 2}\left(1 - e^{-\nu_{\rm m}^{\rm D} z}\right)\mathbb{E}\left[e^{-zI_{\rm u}}\right]\mathbb{E}\left[e^{-zI_{\rm oc}^{\rm mu}}\right] dz. \end{aligned} \tag{39}$$

Due to the composition of independent PPPs and the displacement theorem [30], the interference $I_{\rm u}$ follows a homogeneous PPP with spatial density $\tilde{\lambda}_{\rm u} = (1 - \tau_{\rm m})\lambda_{\rm m} K_{\rm m} +$

$(1-\tau_{\rm s})\,\lambda_{\rm s}K_{\rm s}$, and the corresponding Laplace transform is given as [60]

$$\mathbb{E}\left[e^{-zI_{\rm u}}\right]=\exp\left(-\frac{2\pi^2\tilde{\lambda}_{\rm u}\mathbb{E}[S_{\rm D}^{\frac{2}{\alpha}}]P_{\rm ut}^{\frac{2}{\alpha}}z^{\frac{2}{\alpha}}}{\alpha\sin\left(\frac{2\pi}{\alpha}\right)}\right). \quad (40)$$

As for the Laplace transform of $I_{\rm oc}^{\rm mu}$, the conditional Laplace transform on $L_{\hat{x}_{\rm m},o}$ can be computed as

$$\mathbb{E}\left[e^{-zI_{\rm oc}^{\rm mu}}|L_{\hat{x}_{\rm m},o}{=}t\right]=\exp\left(-\tau_{\rm m}a_{\rm m}\mathcal{C}_{\alpha,K_{\rm m}}(zP_{\rm mt},t)\left(\frac{zP_{\rm mt}}{K_{\rm m}}\right)^{\delta}\right.$$
$$\left.-\,\tau_{\rm s}a_{\rm s}\mathcal{C}_{\alpha,K_{\rm s}}(zP_{\rm mt},t)\left(\frac{zP_{\rm st}}{K_{\rm s}}\right)^{\delta}\right). \quad (41)$$

Notice that $L_{\hat{x}_{\rm m},o}$ has its distribution given by (13), and the rate $R_{\rm m}^{\rm DL}$ given as

$$R_{\rm m}^{\rm DL}=(1-\zeta_{\rm b})\,\mathbb{E}\left[\log_2\left(1+\bar{\gamma}_{\rm m}^{\rm DL}\right)\right], \quad (42)$$

substituting (40) and (41) into (39), and decondition $L_{\hat{x}_{\rm m},o}$ with respect to (13) we have the corresponding result.

### B. Proof of Lemma 3

Let us consider a UE transmitting in uplink to a typical MBS located at the origin, which employs a ZF receive filter $\mathbf{r}_{o,\hat{x}_{\rm u}}^{*}=\hat{\mathbf{h}}_{o,\hat{x}_{\rm u}}^{*}(\sum_{u\in\mathcal{U}_o}\hat{\mathbf{h}}_{o,u}\hat{\mathbf{h}}_{o,u}^{*})^{-1}$ [62], the SINR is then given by

$$\gamma_{\rm m}^{\rm UL}=\frac{P_{\rm ut}L_{o,\hat{x}_{\rm u}}^{-1}|\mathbf{r}_{o,\hat{x}_{\rm u}}^{*}\mathbf{h}_{o,\hat{x}_{\rm u}}|^2}{\left(I_{\rm oc}^{\rm mbs}+I_{\rm u}+\sigma^2\right)\|\mathbf{r}_{o,\hat{x}_{\rm u}}\|^2} \quad (43)$$

where $I_{\rm oc}^{\rm mbs}$ denotes the interference from other cells received at the MBS. By conditioning on the interference, when $K_{\rm m},M_{\rm m}\to\infty$ with $\beta_{\rm m}=K_{\rm m}/M_{\rm m}<1$, the SINR above converges to [45]

$$\gamma_{\rm m}^{\rm UL}\to\bar{\gamma}_{\rm m}^{\rm UL}=\frac{P_{\rm ut}M_{\rm m}(1-\beta_{\rm m})L_{o,\hat{x}_{\rm u}}^{-1}}{I_{\rm oc}^{\rm mbs}+I_{\rm u}+\sigma^2},\quad a.s. \quad (44)$$

By using the continuous mapping theorem [64], the uplink ergodic rate can be calculated as

$$\mathbb{E}\left[\log_2\left(1+\bar{\gamma}_{\rm m}^{\rm UL}\right)\right]=\frac{1}{\ln 2}\mathbb{E}\left[\ln\left(1+\frac{\nu_{\rm m}^{\rm U}L_{o,x_{\rm u}}^{-1}}{I_{\rm oc}^{\rm mbs}+I_{\rm u}+\sigma^2}\right)\right]$$
$$=\int_0^\infty\!\!\int_0^\infty\frac{\left(1-e^{-z\nu_{\rm m}^{\rm U}/t}\right)}{ze^{\sigma^2z}\ln 2}\mathbb{E}\left[e^{-zI_{\rm u}}\right]\mathbb{E}\left[e^{-zI_{\rm oc}^{\rm mbs}}\right]f_{L_{\rm m}}(t)dzdt. \quad (45)$$

The Laplace transform of $I_{\rm oc}^{\rm mbs}$ can be computed as

$$\mathbb{E}\left[e^{-zI_{\rm oc}^{\rm mbs}}\right]$$
$$=\exp\left(-\frac{\Gamma(1{+}\delta)\,\delta\pi^2z^{\delta}}{\sin(\delta\pi)}\left[\frac{\tau_{\rm m}a_{\rm m}P_{\rm mt}^{\delta}\prod_{i=1}^{K_{\rm m}-1}(i{+}\delta)}{\Gamma(K_{\rm m})K_{\rm m}^{\delta}}\right.\right.$$
$$\left.\left.+\;\frac{\tau_{\rm s}a_{\rm s}P_{\rm st}^{\delta}\prod_{i=1}^{K_{\rm s}-1}(i{+}\delta)}{\Gamma(K_{\rm s})K_{\rm s}^{\delta}}\right]\right). \quad (46)$$

On the other hand, to consider the uplink interference from UEs, we use the result in [65] where the path loss from MBS UEs and SAP UEs are modeled as two independent inhomogeneous PPP with intensity measure being

$$\Lambda_{\rm mu}^{\rm (m)}(dx)=\delta a_{\rm m}x^{\delta-1}\left[1-\exp\left(-G_{\rm m}x^{\delta}\right)\right], \quad (47)$$
$$\Lambda_{\rm su}^{\rm (m)}(dx)=\delta a_{\rm s}x^{\delta-1}\left[1-\exp\left(-G_{\rm m}x^{\delta}\right)\right]. \quad (48)$$

The Laplace transform of the UE interference can then be calculated as

$$\mathbb{E}[e^{-zI_{\rm u}}]=\exp\left(-(1-\tau_{\rm m})K_{\rm m}\int_0^\infty\frac{\Lambda_{\rm mu}^{\rm (m)}(dx)}{1+z^{-1}x/P_{\rm ut}}\right.$$
$$\left.-\,(1-\tau_{\rm s})K_{\rm s}\int_0^\infty\frac{\Lambda_{\rm su}^{\rm (m)}(dx)}{1+z^{-1}x/P_{\rm ut}}\right)$$
$$=\exp\left(-\tilde{\lambda}_{\rm u}\pi\mathbb{E}[S_{\rm D}^{\delta}]\int_0^\infty\frac{1-e^{-G_{\rm m}u}}{1+z^{-1}u^{\frac{1}{\delta}}/P_{\rm ut}}du\right) \quad (49)$$

As such, noticing that

$$R_{\rm m}^{\rm UL}=(1-\zeta_{\rm b})\,\mathbb{E}\left[\log_2\left(1+\bar{\gamma}_{\rm m}^{\rm UL}\right)\right] \quad (50)$$

the result follows by substituting (46) and (49) into (45).

### C. Proof of Lemma 4

With a similar approach in the proof of Lemma 2, it is sufficient to consider a typical UE that locates at the origin and associates with an SAP, the corresponding downlink SINR can then be written as

$$\gamma_{\rm s}^{\rm DL}=\frac{P_{\rm st}|\mathbf{h}_{\hat{x}_{\rm s},o}^{*}\mathbf{w}_{\hat{x}_{\rm s},o}|^2L_{\hat{x}_{\rm s},o}^{-1}}{I_{\rm oc}^{\rm su}+I_{\rm u}+\sigma^2} \quad (51)$$

where $\mathbf{w}_{\hat{x}_{\rm s},o}$ is the ZF precoding vector, and $I_{\rm oc}^{\rm su}$ is the aggregate interference from other cells to the SAP UE, given as follows

$$I_{\rm oc}^{\rm su}=\sum_{x_{\rm s}\in\Phi_{\rm s}^{\rm D}\backslash\hat{x}_{\rm s}}\frac{P_{\rm st}g_{x_{\rm s},o}}{K_{\rm s}L_{x_{\rm s},o}}+\sum_{x_{\rm m}\in\Phi_{\rm m}^{\rm D}}\frac{P_{\rm mt}g_{x_{\rm m},o}}{K_{\rm m}L_{x_{\rm m},o}}. \quad (52)$$

By noting that $|K_{\rm s}\mathbf{h}_{\hat{x}_{\rm s},o}^{*}\mathbf{w}_{\hat{x}_{\rm s},o}|^2\sim\Gamma(\Delta_{\rm s},1)$ with $\Delta_{\rm s}=M_{\rm s}-K_{\rm s}+1$ [14], we can write the rate of an SAP UE as [64]

$$\mathbb{E}\left[\log_2\left(1+\gamma_{\rm s}^{\rm DL}\right)\right]=\int_0^\infty\!\!\int_0^\infty\frac{e^{-\sigma^2z}}{z\ln 2}\mathbb{E}\left[e^{-zI_{\rm oc}^{\rm su}}\right]\mathbb{E}\left[e^{-zI_{\rm u}}\right]$$
$$\times\left[1-\frac{1}{(1{+}zP_{\rm st}t^{-1}/K_{\rm s})^{\Delta_{\rm s}}}\right]f_{L_{\rm s}}(t)dzdt. \quad (53)$$

On one hand, $\mathbb{E}\left[e^{-zI_{\rm u}}\right]$ is given in (40), on the other, by denoting $L_{x_{\rm s},o}=t$, the conditional Laplace transform of $I_{\rm oc}^{\rm su}$ can be derived as

$$\mathbb{E}\left[e^{-zI_{\rm oc}^{\rm su}}|L_{\hat{x}_{\rm s},o}=t\right]=\exp\left(-\tau_{\rm s}a_{\rm s}\mathcal{C}_{\alpha,K_{\rm s}}(zP_{\rm st},t)\left(\frac{zP_{\rm st}}{K_{\rm s}}\right)^{\delta}\right.$$
$$\left.-\,\tau_{\rm m}a_{\rm m}\mathcal{C}_{\alpha,K_{\rm m}}(zP_{\rm st},t)\left(\frac{zP_{\rm mt}}{K_{\rm m}}\right)^{\delta}\right). \quad (54)$$

As $L_{\hat{x}_{\rm s},o}$ follows a distribution as (17), and $R_{\rm s}^{\rm DL}$ is given as

$$R_{\rm s}^{\rm DL}=(1-\zeta_{\rm b})\,\mathbb{E}\left[\log_2\left(1+\gamma_{\rm s}^{\rm DL}\right)\right] \quad (55)$$

by substituting (40) and (54) into (53), and deconditioning $L_{\hat{x}_{\rm s},o}$, we have the desired result.

### D. Proof of Lemma 6

We consider the signal received at the $k$-th antenna of a typical SAP located at the origin, which is served by a MBS through the wireless backhaul and has the SINR given as

$$\gamma_{\mathrm{b},k}^{\mathrm{DL}} = \frac{P_{\mathrm{mb}} \left| \mathbf{h}_{b,k}^{*} \mathbf{v}_{b,k} \right|^{2} L_{\hat{x}_{\mathrm{b}},o}^{-1}}{\sigma^{2} + I_{\mathrm{m}} + I_{\mathrm{s}}} \tag{56}$$

where $\mathbf{v}_{b,k}$ denotes the ZF precoder, while $I_{\mathrm{m}}$ and $I_{\mathrm{s}}$ are the aggregated interference from downlink transmitting MBSs and uplink transmitting SAPs in the wireless backhaul, respectively, with expression given as follows

$$I_{\mathrm{m}} = \sum_{x_{\mathrm{m}} \in \Phi_{\mathrm{m}}^{\mathrm{D}} \backslash \hat{x}_{\mathrm{m}}} \frac{P_{\mathrm{mb}} \tilde{g}_{x_{\mathrm{m}},o}}{K_{\mathrm{b}} M_{\mathrm{s}} L_{x_{\mathrm{m}},o}} \tag{57}$$

and

$$I_{\mathrm{s}} = \sum_{x_{\mathrm{s}} \in \Phi_{\mathrm{s}}^{\mathrm{U}}} \frac{P_{\mathrm{sb}} \tilde{g}_{x_{\mathrm{s}},o}}{M_{\mathrm{s}} L_{x_{\mathrm{s}},o}} \tag{58}$$

where $\tilde{g}_{x_{\mathrm{m}},o} \sim \Gamma(K_{\mathrm{b}} M_{\mathrm{s}}, 1)$ and $\tilde{g}_{x_{\mathrm{s}},o} \sim \Gamma(M_{\mathrm{s}}, 1)$ are the effective small scale fading from the interfering MBS $x_{\mathrm{m}}$ and SAP $x_{\mathrm{s}}$ to the origin.

By conditioning on the interference $I_{\mathrm{m}} + I_{\mathrm{s}}$, and by using a similar approach as the one in the proof of Lemma 2, when $K_{\mathrm{b}}, M_{\mathrm{m}} \to \infty$ with $\beta_{\mathrm{b}} = K_{\mathrm{b}} M_{\mathrm{s}} / M_{\mathrm{m}} < 1$, the SINR in (56) satisfies [45]

$$\gamma_{\mathrm{b},k}^{\mathrm{DL}} \to \bar{\gamma}_{\mathrm{b},k}^{\mathrm{DL}} = \frac{P_{\mathrm{mb}} (1 - \beta_{\mathrm{b}}) a_{\mathrm{b}}^{\delta}}{\beta_{\mathrm{b}} \Gamma\left(1 + \frac{1}{\delta}\right) (\sigma^{2} + I_{\mathrm{m}} + I_{\mathrm{s}})}, \quad a.s. \tag{59}$$

By using the continuous mapping theorem and the lemma in [64], we have the following holds

$$\mathbb{E}\left[\log_{2}\left(1 + \bar{\gamma}_{\mathrm{b},k}^{\mathrm{DL}}\right)\right] = \int_{0}^{\infty} \frac{e^{-\sigma^{2} z}}{z \ln 2} \left(1 - e^{-z \nu_{\mathrm{b}}^{\mathrm{D}}}\right) \times \mathbb{E}\left[e^{-z I_{\mathrm{s}}}\right] \mathbb{E}\left[e^{-z I_{\mathrm{m}}}\right] dz. \tag{60}$$

With the effective channel distribution available, we can compute the Laplace transform of $I_{\mathrm{s}}$ as

$$\mathbb{E}[e^{-z I_{\mathrm{s}}}] = \exp\left(-\frac{\Gamma(1+\delta) \delta \pi^{2} z^{\delta} P_{\mathrm{sb}}^{\delta}}{\sin(\delta\pi) \Gamma(M_{\mathrm{s}}) M_{\mathrm{s}}^{\delta}} \mathbb{E}[S_{\mathrm{B}}^{\delta}] (1 - \tau_{\mathrm{b}}) \lambda_{\mathrm{s}} \prod_{i=1}^{M_{\mathrm{s}}-1} (i + \delta)\right). \tag{61}$$

On the other hand, conditioning on $L_{\hat{x}_{\mathrm{b}},o} = t$, we have the conditional Laplace transform given as

$$\mathbb{E}\left[e^{-z I_{\mathrm{m}}} | L_{\hat{x}_{\mathrm{b}},o} = t\right] = \exp\left(-\tau_{\mathrm{b}} a_{\mathrm{b}} \mathcal{C}_{\alpha, K_{\mathrm{b}} M_{\mathrm{s}}}\left(z P_{\mathrm{mb}}, t\right) \left(\frac{z P_{\mathrm{mb}}}{K_{\mathrm{b}} M_{\mathrm{s}}}\right)^{\delta}\right). \tag{62}$$

Notice that $L_{\hat{x}_{\mathrm{b}},o}$ has its distribution as (20), and the downlink rate achievable on the wireless backhaul given as

$$R_{\mathrm{b}}^{\mathrm{DL}} = \frac{\zeta_{\mathrm{b}} M_{\mathrm{s}}}{K_{\mathrm{s}}} \mathbb{E}\left[\log_{2}\left(1 + \bar{\gamma}_{\mathrm{b},k}^{\mathrm{DL}}\right)\right] \tag{63}$$

Lemma 6 then follows from substituting (61) and (62) into (60), and deconditioning with respect to (20).

### E. Proof of Lemma 7

Let us consider an SAP transmitting in uplink to a typical MBS located at the origin, which employs a ZF receive filter, the received SINR from the $k^{\mathrm{th}}$ antenna is then given by

$$\gamma_{\mathrm{b},k}^{\mathrm{UL}} = \frac{P_{\mathrm{sb}} L_{o,\hat{x}_{\mathrm{b}}}^{-1} |\mathbf{r}_{\mathrm{b},k}^{*} \mathbf{h}_{\mathrm{b},k}|^{2}}{(I_{\mathrm{m}} + I_{\mathrm{s}} + \sigma^{2}) \|\mathbf{r}_{\mathrm{b},k}\|^{2}} \tag{64}$$

By conditioning on the interference $I_{\mathrm{m}} + I_{\mathrm{s}}$, when $K_{\mathrm{b}}, M_{\mathrm{m}} \to \infty$ with $\beta_{\mathrm{b}} = K_{\mathrm{b}} M_{\mathrm{s}} / M_{\mathrm{m}} < 1$, the SINR $\gamma_{\mathrm{b},k}^{\mathrm{UL}}$ satisfies [45]

$$\gamma_{\mathrm{b},k}^{\mathrm{UL}} \to \bar{\gamma}_{\mathrm{b},k}^{\mathrm{DL}} = \frac{P_{\mathrm{sb}} M_{\mathrm{m}} (1 - \beta_{\mathrm{b}}) L_{\hat{x}_{\mathrm{b}},o}^{-1}}{\sigma^{2} + I_{\mathrm{m}} + I_{\mathrm{s}}}, \quad a.s. \tag{65}$$

By using the continuous mapping theorem and the lemma in [64], we have the following holds

$$\mathbb{E}\left[\log_{2}\left(1 + \bar{\gamma}_{\mathrm{b},k}^{\mathrm{UL}}\right)\right] = \int_{0}^{\infty} \frac{e^{-\sigma^{2} z}}{z \ln 2} \left(1 - e^{-z \nu_{\mathrm{b}}^{\mathrm{U}}}\right) \times \mathbb{E}\left[e^{-z I_{\mathrm{s}}}\right] \mathbb{E}\left[e^{-z I_{\mathrm{m}}}\right] dz. \tag{66}$$

On one hand, the Laplace transform of $I_{\mathrm{m}}$ can be calculated as

$$\mathbb{E}\left[e^{-z I_{\mathrm{m}}}\right] = \exp\left\{-\frac{\tau_{\mathrm{b}} a_{\mathrm{b}} \Gamma(1+\delta) \delta \pi^{2} P_{\mathrm{mb}}^{\delta} z^{\delta} \prod_{i=1}^{K_{\mathrm{b}} M_{\mathrm{s}} - 1} (i + \delta)}{\sin(\delta\pi) \left(M_{\mathrm{s}} K_{\mathrm{b}}\right)^{\delta} \Gamma\left(M_{\mathrm{s}} K_{\mathrm{b}}\right)}\right\}. \tag{67}$$

On the other, to consider the uplink interference from SAPs, we use the result in [65] where the path loss from the interfering SAPs are modeled as an inhomogeneous PPP with intensity measure being

$$\Lambda_{\mathrm{s}}^{(\mathrm{m})}(dx) = \delta a_{\mathrm{b}} x^{\delta - 1} \left[1 - \exp\left(-a_{\mathrm{b}} x^{\delta}\right)\right]. \tag{68}$$

As such, the Laplace transform of $I_{\mathrm{s}}$ in the uplink can be computed as

$$\mathbb{E}\left[e^{-z I_{\mathrm{s}}}\right] = \exp\left(-\left(1 - \tau_{\mathrm{b}}\right) K_{\mathrm{b}} \times \int_{0}^{\infty} \left[1 - \frac{1}{\left(1 + z P_{\mathrm{sb}} x^{-1} / M_{\mathrm{s}}\right)^{M_{\mathrm{s}}}}\right] \Lambda_{\mathrm{s}}^{(\mathrm{m})}(dx)\right). \tag{69}$$

Since the uplink rate achievable on the wireless backhaul is given as

$$R_{\mathrm{b}}^{\mathrm{UL}} = \frac{\zeta_{\mathrm{b}} M_{\mathrm{s}}}{K_{\mathrm{s}}} \mathbb{E}\left[\log_{2}\left(1 + \bar{\gamma}_{\mathrm{b},k}^{\mathrm{UL}}\right)\right] \tag{70}$$

the result then follows by substituting (67) and (69) into (66).

### F. Proof of Lemma 8

The average rate for a typical UE located at the origin is given by

$$R = A_{\mathrm{m}} R_{\mathrm{m}} + A_{\mathrm{s}} R_{\mathrm{s}} \tag{71}$$

where $R_{\mathrm{m}}$ and $R_{\mathrm{s}}$ are the data rates when the UE associates to a MBS and a SAP, respectively, given by

$$R_{\mathrm{m}} = \tau_{\mathrm{m}} R_{\mathrm{m}}^{\mathrm{DL}} + (1 - \tau_{\mathrm{m}}) R_{\mathrm{m}}^{\mathrm{UL}} \tag{72}$$

and

$$R_{\rm s}=\tau_{\rm s}\min\left\{R_{\rm s}^{\rm DL},R_{\rm b}^{\rm DL}\right\}+(1-\tau_{\rm s})\min\left\{R_{\rm s}^{\rm UL},R_{\rm b}^{\rm UL}\right\}. \quad (73)$$

As each MBS and each SAP serve $K_{\rm m}$ and $K_{\rm s}$ UEs, respectively, the total density of active UEs is given by $K_{\rm m}\lambda_{\rm m}+K_{\rm s}\lambda_{\rm s}$. Let $B$ be the available bandwidth, the sum rate per area is obtained as $\mathcal{R}=(K_{\rm m}\lambda_{\rm m}+K_{\rm s}\lambda_{\rm s})\,BR$. Lemma 8 then follows from Lemmas 2 to 7 and by the continuous mapping theorem.

### *G. Proof of Lemma 9*

We consider a typical UE that locates at the origin, notice that under FDD, the wireless link experiences interference from the downlink transmitting MBSs and SAPs. As such, the SINR can be written as

$$\gamma_{\rm m,FDD}^{\rm DL}=\frac{P_{\rm mt}|\mathbf{h}_{\hat{x}_{\rm m},o}^{*}\mathbf{w}_{\hat{x}_{\rm m},o}|^{2}L_{\hat{x}_{\rm m},o}^{-1}}{I_{\rm oc}^{\rm mu}+\sigma^{2}}. \quad (74)$$

By conditioning on the interference, when $K_{\rm m},M_{\rm m}\to\infty$ with $\beta_{\rm m}=K_{\rm m}/M_{\rm m}<1$, the SINR under ZF precoding converges to [45]

$$\gamma_{\rm m,FDD}^{\rm DL}\to\bar{\gamma}_{\rm m,FDD}^{\rm DL}=\frac{\nu_{\rm m}^{\rm D}}{I_{\rm oc}^{\rm mu}+\sigma^{2}},\ \ a.s. \quad (75)$$

by using the continuous mapping theorem, and the lemma in [64], we have the following

$$\mathbb{E}\left[\log_2\left(1+\bar{\gamma}_{\rm m,FDD}^{\rm DL}\right)\right]=\int_0^\infty\frac{\left(1-e^{-z\nu_{\rm m}^{\rm D}}\right)}{ze^{\sigma^2 z}\ln 2}\mathbb{E}\left[e^{-zI_{\rm oc}^{\rm mu}}\right]dz \quad (76)$$

The result then follows by noticing that $R_{\rm m,FDD}^{\rm DL}=\xi_{\rm D}(1-\zeta_{\rm b})\mathbb{E}[\log_2(1+\bar{\gamma}_{\rm m,FDD}^{\rm DL})]$, and by substituting (41) into (76).

## References

[1] H. H. Yang, G. Geraci, and T. Q. S. Quek, "Rate analysis of spatial multiplexing in MIMO heterogeneous networks with wireless backhaul," in *Proc. Internat. Conf. Acoust. Speech Signal Process.*, Shanghai, China, Mar. 2016.

[2] ——, "MIMO hetnets with wireless backhaul: An energy-efficient design," in *Proc. IEEE Int. Conf. on Comm. (ICC)*, Kuala Lumpur, Malaysia, May 2016.

[3] G. Auer, V. Giannini, C. Desset, I. Godor, P. Skillermark, M. Olsson, M. Imran, D. Sabella, M. Gonzalez, O. Blume, and A. Fehske, "How much energy is needed to run a wireless network?" *IEEE Wireless Commun.*, vol. 18, no. 5, pp. 40–49, Oct. 2011.

[4] Y. Chen, S. Zhang, S. Xu, and G. Y. Li, "Fundamental trade-offs on green wireless networks," *IEEE Commun. Mag.*, vol. 49, no. 6, pp. 30–37, Jun. 2011.

[5] G. Y. Li, Z. Xu, C. Xiong, C. Yang, S. Zhang, Y. Chen, and S. Xu, "Energy-efficient wireless communications: Tutorial, survey, and open issues," *IEEE Trans. Wireless Commun.*, vol. 18, no. 6, pp. 28–35, Dec. 2011.

[6] D. Feng, C. Jiang, G. Lim, L. J. Cimini Jr, G. Feng, and G. Y. Li, "A survey of energy-efficient wireless communications," *IEEE Commun. Surveys and Tutorials*, vol. 15, no. 1, pp. 167–178, Feb. 2013.

[7] R. Hu and Y. Qian, "An energy efficient and spectrum efficient wireless heterogeneous network framework for 5G systems," *IEEE Commun. Mag.*, vol. 52, no. 5, pp. 94–101, May 2014.

[8] G. Geraci, M. Wildemeersch, and T. Q. S. Quek, "Energy efficiency of distributed signal processing in wireless networks: A cross-layer analysis," *IEEE Trans. Signal Process.*, vol. 64, no. 4, pp. 1034–1047, Feb. 2016.

[9] H. H. Yang, J. Lee, and T. Q. S. Quek, "Heterogeneous cellular network with energy harvesting-based D2D communication," *IEEE Trans. Wireless Commun.*, vol. 15, no. 2, pp. 1406–1419, Feb. 2016.

[10] T. Q. S. Quek, G. de la Roche, I. Güvenç, and M. Kountouris, *Small cell networks: Deployment, PHY techniques, and resource management*. Cambridge University Press, 2013.

[11] J. G. Andrews, H. Claussen, M. Dohler, S. Rangan, and M. C. Reed, "Femtocells: Past, present, and future," *IEEE J. Sel. Areas Commun.*, vol. 30, no. 3, pp. 497–508, Apr. 2012.

[12] J. Hoydis, M. Kobayashi, and M. Debbah, "Green small-cell networks," *IEEE Vehicular Technology Mag.*, vol. 6, no. 1, pp. 37–43, Mar. 2011.

[13] Q. Ye, B. Rong, Y. Chen, M. Al-Shalash, C. Caramanis, and J. G. Andrews, "User association for load balancing in heterogeneous cellular networks," *IEEE Trans. Wireless Commun.*, vol. 12, no. 6, pp. 2706–2716, Jun. 2013.

[14] H. S. Dhillon, M. Kountouris, and J. G. Andrews, "Downlink MIMO hetnets: Modeling, ordering results and performance analysis," *IEEE Trans. Wireless Commun.*, vol. 12, no. 10, pp. 5208–5222, Oct. 2013.

[15] Small Cell Forum, "Backhaul technologies for small cells," white paper, document 049.05.02, Feb. 2014.

[16] H. S. Dhillon and G. Caire, "Information theoretic upper bound on the capacity of wireless backhaul networks," in *Proc. IEEE Int. Symp. on Inform. Theory*, Honolulu, HI, Jun. 2014, pp. 251–255.

[17] ——, "Scalability of line-of-sight massive MIMO mesh networks for wireless backhaul," in *Proc. IEEE Int. Symp. on Inform. Theory*, Honolulu, HI, Jun. 2014, pp. 2709–2713.

[18] L. Sanguinetti, A. L. Moustakas, and M. Debbah, "Interference management in 5G reverse TDD HetNets: A large system analysis," *IEEE J. Sel. Areas Commun.*, vol. 33, no. 6, pp. 1–1, Mar. 2015.

[19] J. Andrews, "Seven ways that HetNets are a cellular paradigm shift," *IEEE Commun. Mag.*, vol. 51, no. 3, pp. 136–144, Mar. 2013.

[20] H. Claussen, "Future cellular networks," Alcatel-Lucent, Apr. 2012.

[21] A. J. Fehske, F. Richter, and G. P. Fettweis, "Energy efficiency improvements through micro sites in cellular mobile radio networks," in *Proc. IEEE Global Telecomm. Conf. Workshops*, Honolulu, HI, Dec. 2009, pp. 1–5.

[22] C. Li, J. Zhang, and K. Letaief, "Throughput and energy efficiency analysis of small cell networks with multi-antenna base stations," *IEEE Trans. Wireless Commun.*, vol. 13, no. 5, pp. 2505–2517, May 2014.

[23] M. Wildemeersch, T. Q. S. Quek, C. H. Slump, and A. Rabbachin, "Cognitive small cell networks: Energy efficiency and trade-offs," *IEEE Trans. Commun.*, vol. 61, no. 9, pp. 4016–4029, Sep. 2013.

[24] Y. S. Soh, T. Q. S. Quek, M. Kountouris, and H. Shin, "Energy efficient heterogeneous cellular networks," *IEEE J. Sel. Areas Commun.*, vol. 31, no. 5, pp. 840–850, Apr. 2013.

[25] S. Navaratnarajah, A. Saeed, M. Dianati, and M. A. Imran, "Energy efficiency in heterogeneous wireless access networks," *IEEE Wireless Commun.*, vol. 20, no. 5, pp. 37–43, Oct. 2013.

[26] E. Björnson, M. Kountouris, and M. Debbah, "Massive MIMO and small cells: Improving energy efficiency by optimal soft-cell coordination," in *Proc. IEEE Int. Conf. on Telecommun.*, Casablanca, Morocco, May 2013, pp. 1–5.

[27] X. Ge, H. Cheng, M. Guizani, and T. Han, "5G wireless backhaul networks: Challenges and research advances," *IEEE Network*, vol. 28, no. 6, pp. 6–11, Dec. 2014.

[28] S. Tombaz, P. Monti, F. Farias, M. Fiorani, L. Wosinska, and J. Zander, "Is backhaul becoming a bottleneck for green wireless access networks?" in *Proc. IEEE Int. Conf. on Comm. (ICC)*, Sydney, Australia, Jun. 2014, pp. 4029–4035.

[29] S. Tombaz, P. Monti, K. Wang, A. Vastberg, M. Forzati, and J. Zander, "Impact of backhauling power consumption on the deployment of heterogeneous mobile networks," in *Proc. IEEE Global Telecomm. Conf.*, Houston, TX, Dec. 2011, pp. 1–5.

[30] M. Haenggi, *Stochastic geometry for wireless networks*. Cambridge University Press, 2012.

[31] R. Couillet and M. Debbah, *Random matrix methods for wireless communications*. Cambridge University Press, 2011.

[32] D. B. Taylor, H. S. Dhillon, T. D. Novlan, and J. G. Andrews, "Pairwise interaction processes for modeling cellular network topology," in *Proc. IEEE Global Telecomm. Conf.*, Anaheim, CA, Dec. 2012, pp. 4524–4529.

[33] B. Blaszczyszyn, M. K. Karray, and H. P. Keeler, "Using Poisson processes to model lattice cellular networks," in *Proc. IEEE Conf. on Computer Commun.*, Turin, Italy, Apr. 2013, pp. 773–781.

[34] M. Sanchez-Fernandez, S. Zazo, and R. Valenzuela, "Performance comparison between beamforming and spatial multiplexing for the downlink

in wireless cellular systems," *IEEE Trans. Wireless Commun.*, vol. 6, no. 7, pp. 2427–2431, Jul. 2007.

[35] H. Shirani-Mehr, G. Caire, and M. J. Neely, "MIMO downlink scheduling with non-perfect channel state knowledge," *IEEE Trans. Commun.*, vol. 58, no. 7, pp. 2055–2066, Jul. 2010.

[36] V. Chandrasekhar and J. G. Andrews, "Spectrum allocation in tiered cellular networks," *IEEE Trans. Commun.*, vol. 57, no. 10, pp. 3059–3068, Oct. 2009.

[37] W. C. Cheung, T. Q. S. Quek, and M. Kountouris, "Throughput optimization, spectrum allocation, and access control in two-tier femtocell networks," *IEEE J. Sel. Areas Commun.*, vol. 30, no. 3, pp. 561–574, Apr. 2012.

[38] T. Zahir, K. Arshad, A. Nakata, and K. Moessner, "Interference management in femtocells," *IEEE Commun. Surveys and Tutorials*, vol. 15, no. 1, pp. 293–311, 2013.

[39] M. Peng, C. Wang, J. Li, H. Xiang, and V. Lau, "Recent advances in underlay heterogeneous networks: Interference control, resource allocation, and self-organization," *IEEE Commun. Surveys and Tutorials*, vol. 17, no. 2, pp. 700–729, May. 2015.

[40] H. S. Dhillon and G. Caire, "Wireless backhaul networks: Capacity bound, scalability analysis and design guidelines," *IEEE Trans. Wireless Commun.*, vol. 14, no. 11, pp. 6043 – 6056, Nov. 2015.

[41] S. Singh, M. N. Kulkarni, A. Ghosh, and J. G. Andrews, "Tractable model for rate in self-backhauled millimeter wave cellular networks," *IEEE J. Sel. Areas Commun.*, vol. 33, no. 10, pp. 2196–2211, Oct. 2015.

[42] Z. Shen, A. Khoryaev, E. Eriksson, and X. Pan, "Dynamic uplink-downlink configuration and interference management in TD-LTE," *IEEE Commun. Mag.*, vol. 50, no. 11, pp. 51–59, Nov. 2012.

[43] M. Ding, D. Lopez Perez, A. V. Vasilakos, and W. Chen, "Dynamic TDD transmissions in homogeneous small cell networks," in *Proc. IEEE Int. Conf. on Comm. (ICC)*, Sydney, Australia, Jun. 2014, pp. 616–621.

[44] Q. H. Spencer, C. B. Peel, A. L. Swindlehurst, and M. Haardt, "An introduction to the multi-user MIMO downlink," *IEEE Commun. Mag.*, vol. 42, no. 10, pp. 60–67, Oct. 2004.

[45] S. Wagner, R. Couillet, M. Debbah, and D. T. Slock, "Large system analysis of linear precoding in correlated miso broadcast channels under limited feedback," *IEEE Trans. Inf. Theory*, vol. 58, no. 7, pp. 4509–4537, Jul. 2012.

[46] G. Geraci, M. Egan, J. Yuan, A. Razi, and I. B. Collings, "Secrecy sum-rates for multi-user MIMO regularized channel inversion precoding," *IEEE Trans. Commun.*, vol. 60, no. 11, pp. 3472–3482, Nov. 2012.

[47] G. Geraci, A. Y. Al-Nahari, J. Yuan, and I. B. Collings, "Linear precoding for broadcast channels with confidential messages under transmit-side channel correlation," *IEEE Commun. Lett.*, vol. 17, no. 6, pp. 1164–1167, Jun. 2013.

[48] G. Geraci, R. Couillet, J. Yuan, M. Debbah, and I. B. Collings, "Large system analysis of linear precoding in MISO broadcast channels with confidential messages," *IEEE J. Sel. Areas Commun.*, vol. 31, no. 9, pp. 1660–1671, Sep. 2013.

[49] E. Björnson, L. Sanguinetti, J. Hoydis, and M. Debbah, "Optimal design of energy-efficient multi-user MIMO systems: Is massive MIMO the answer?" *IEEE Trans. Wireless Commun.*, vol. 14, no. 6, pp. 3059 – 3075, Jun. 2015.

[50] A. Mezghani and J. A. Nossek, "Power efficiency in communication systems from a circuit perspective," in *IEEE Int. Symp. on Circuits and Systems*, Rio de Janeiro, May 2011, pp. 1896–1899.

[51] S. Tombaz, A. Västberg, and J. Zander, "Energy-and cost-efficient ultra-high-capacity wireless access," *IEEE Wireless Commun.*, vol. 18, no. 5, pp. 18–24, Oct. 2011.

[52] H. Yang and T. L. Marzetta, "Total energy efficiency of cellular large scale antenna system multiple access mobile networks," in *IEEE Online GreenCom*, Oct. 2013, pp. 27–32.

[53] T. Chen, H. Kim, and Y. Yang, "Energy efficiency metrics for green wireless communications," in *Proc. IEEE Int. Conf. Wireless Commun. and Signal Processing*, Suzhou, China, 2010, pp. 1–6.

[54] S. Singh, H. S. Dhillon, and J. G. Andrews, "Offloading in heterogeneous networks: Modeling, analysis, and design insights," *IEEE Trans. Wireless Commun.*, vol. 12, no. 5, pp. 2484–2497, May 2013.

[55] G. Caire and S. Shamai, "On the achievable throughput of a multiantenna gaussian broadcast channel," *IEEE Trans. Inf. Theory*, vol. 49, no. 7, pp. 1691–1706, Jul. 2003.

[56] Q. H. Spencer, A. L. Swindlehurst, and M. Haardt, "Zero-forcing methods for downlink spatial multiplexing in multiuser MIMO channels," *IEEE Trans. Signal Process.*, vol. 52, no. 2, pp. 461–471, Feb. 2004.

[57] T. Yoo and A. Goldsmith, "On the optimality of multiantenna broadcast scheduling using zero-forcing beamforming," *IEEE J. Sel. Areas Commun.*, vol. 24, no. 3, pp. 528–541, Mar. 2006.

[58] G. Auer, V. Giannini, C. Desset, I. Godor, P. Skillermark, M. Olsson, M. A. Imran, D. Sabella, M. J. Gonzalez, O. Blume *et al.*, "How much energy is needed to run a wireless network?" *IEEE Wireless Commun. Mag.*, vol. 18, no. 5, pp. 40–49, Oct. 2011.

[59] B. Debaillie, C. Desset, and F. Louagie, "A flexible and future-proof power model for cellular base stations," in *IEEE Vehicular Technology Conference*, Glasgow, Scotland, May 2015, pp. 1–7.

[60] F. Baccelli and B. Blaszczyszyn, *Stochastic Geometry and Wireless Networks. Volumn I: Theory*. Now Publishers, 2009.

[61] A. J. Goldsmith and S. B. Wicker, "Design challenges for energy-constrained ad hoc wireless networks," *IEEE Wireless Communications*, vol. 9, no. 4, pp. 8–27, Aug. 2002.

[62] D. N. C. Tse and P. Viswanath, *Fundamentals of Wireless Communication*. Cambridge University Press, 2005.

[63] G. Geraci, H. S. Dhillon, J. G. Andrews, J. Yuan, and I. Collings, "Physical layer security in downlink multi-antenna cellular networks," *IEEE Trans. Commun.*, vol. 62, no. 6, pp. 2006–2021, Jun. 2014.

[64] K. A. Hamdi, "A useful lemma for capacity analysis of fading interference channels," *IEEE Trans. Commun.*, vol. 58, no. 2, pp. 411–416, Feb. 2010.

[65] S. Singh, X. Zhang, and J. Andrews, "Joint rate and SINR coverage analysis for decoupled uplink-downlink biased cell associations in hetnets," *IEEE Trans. Wireless Commun.*, vol. 14, no. 10, pp. 5360–5373, Oct. 2015.

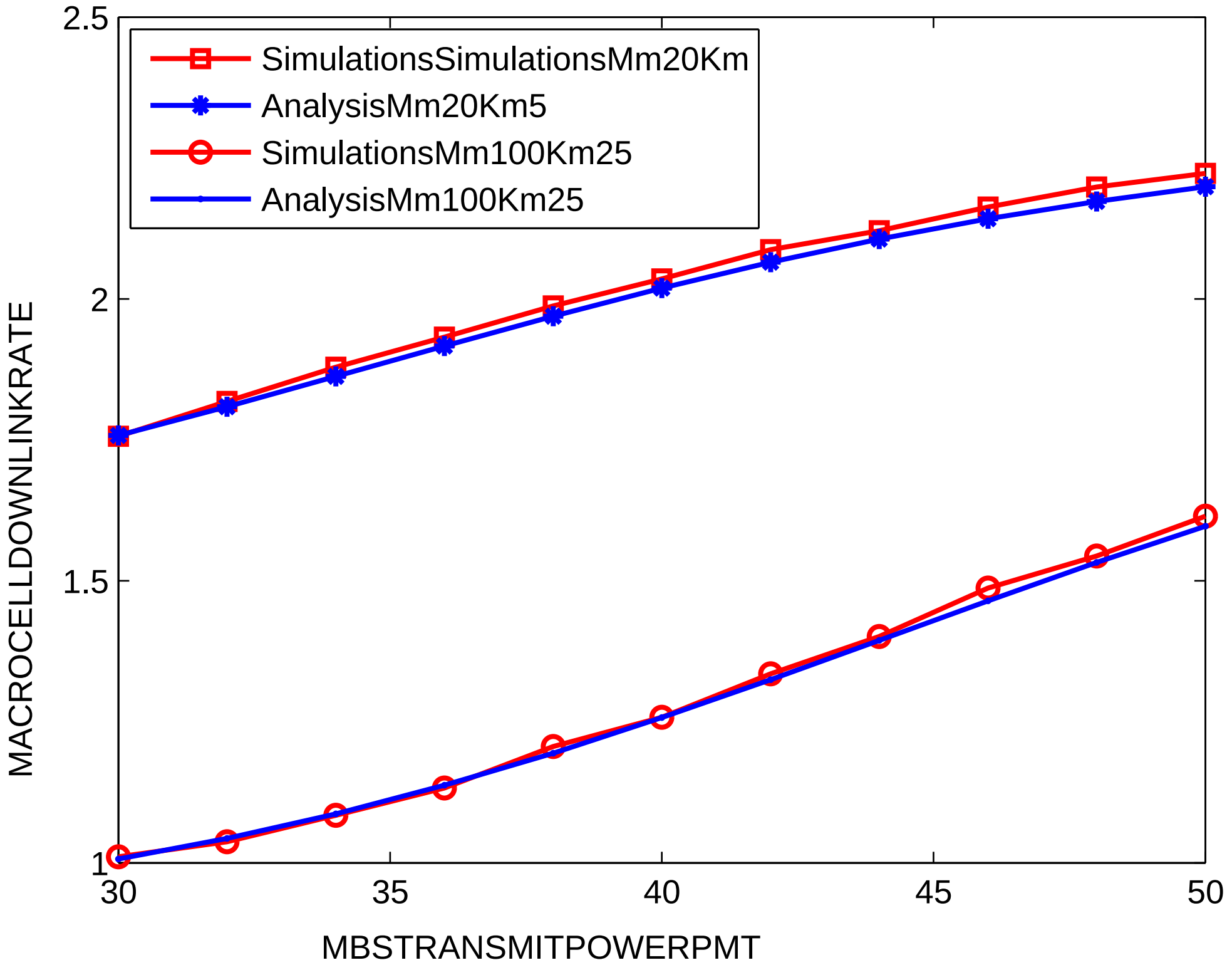

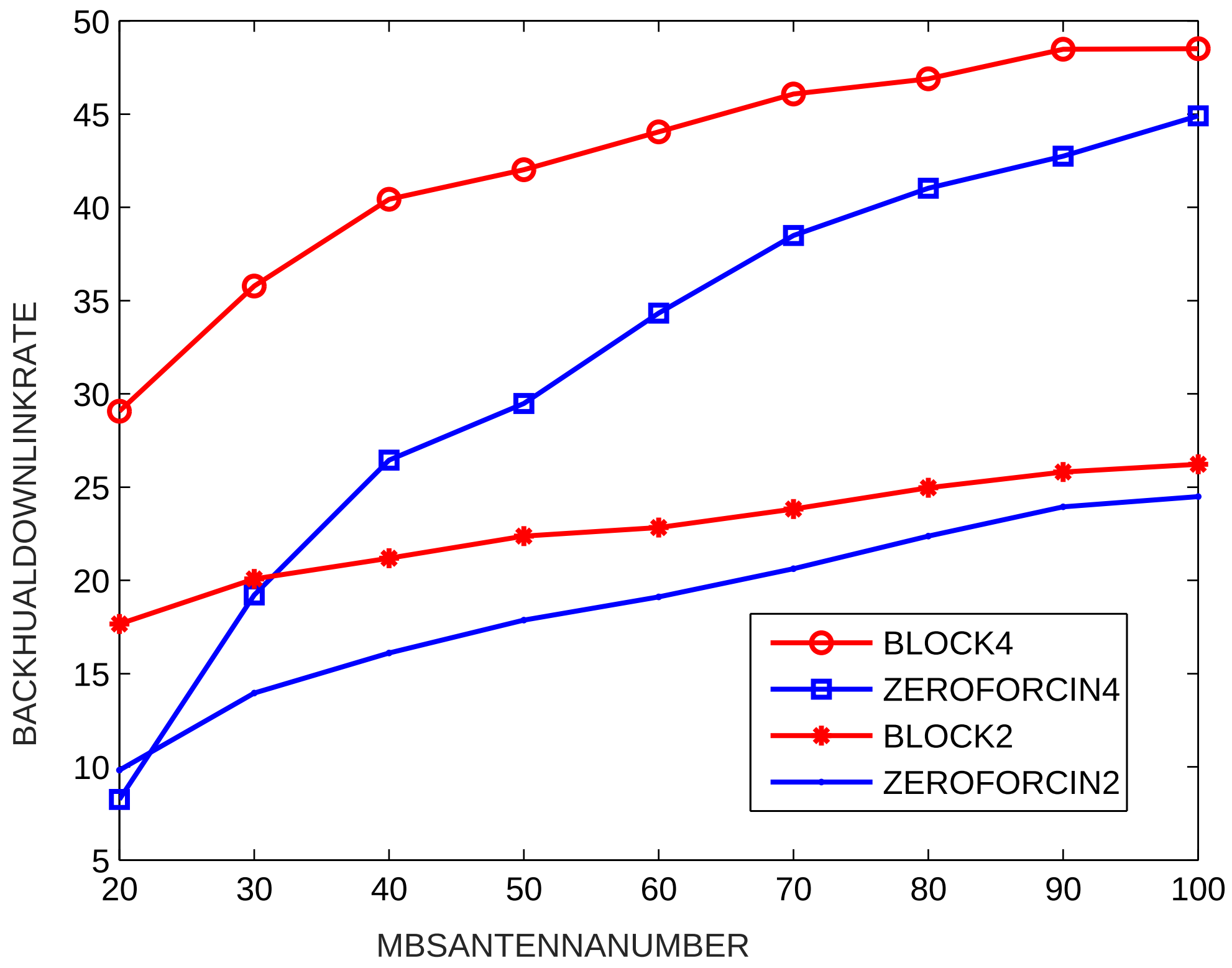

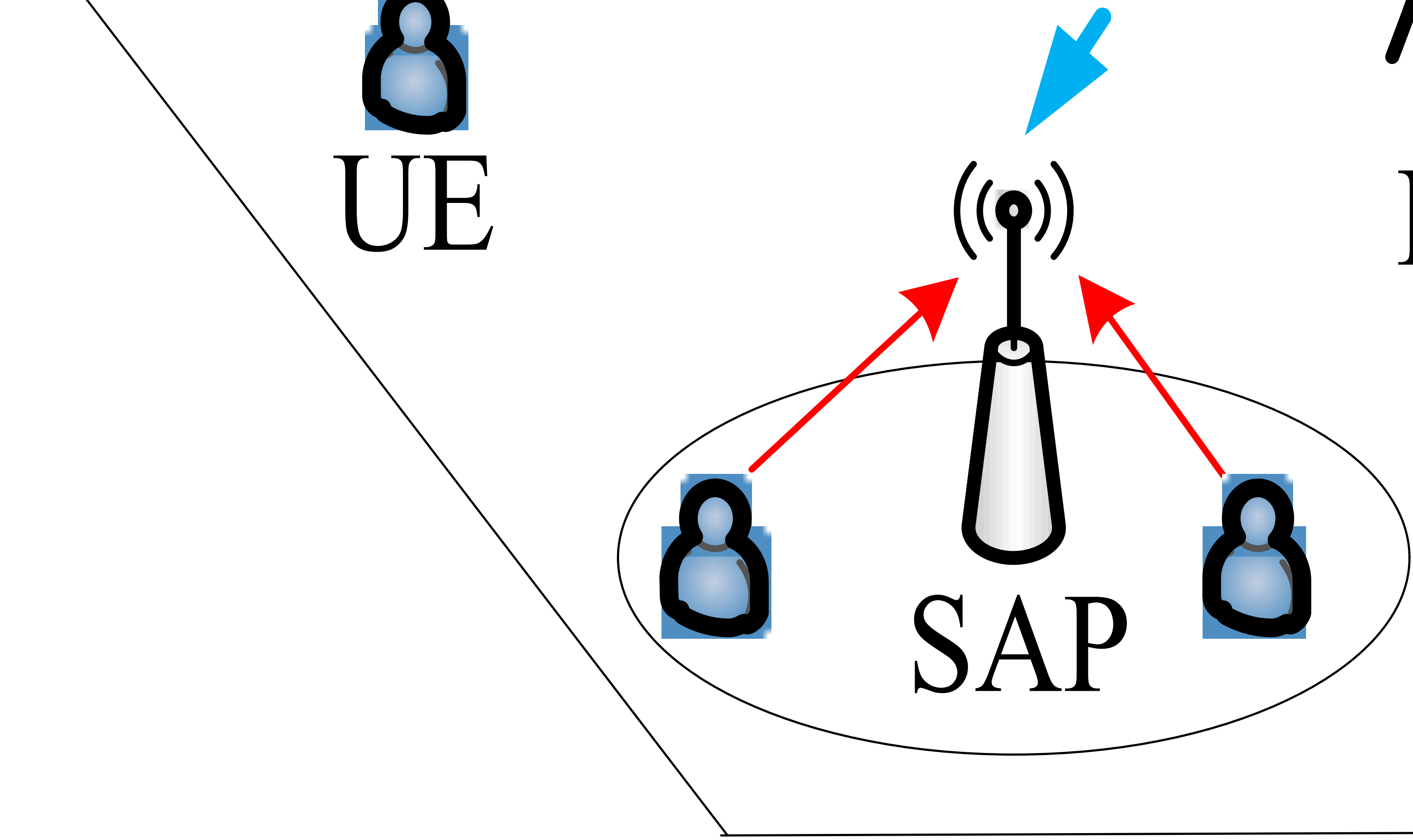
UE
SAP

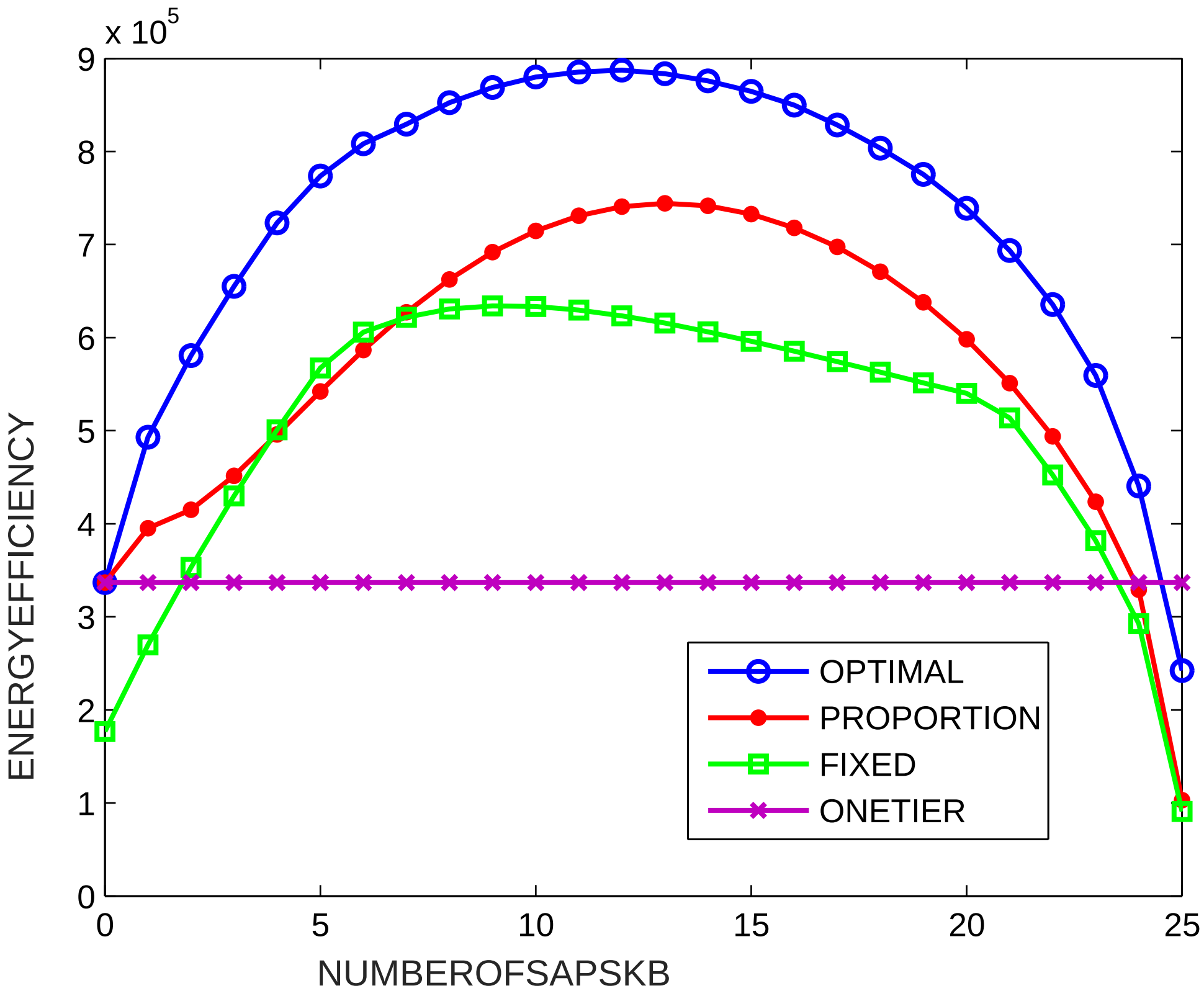

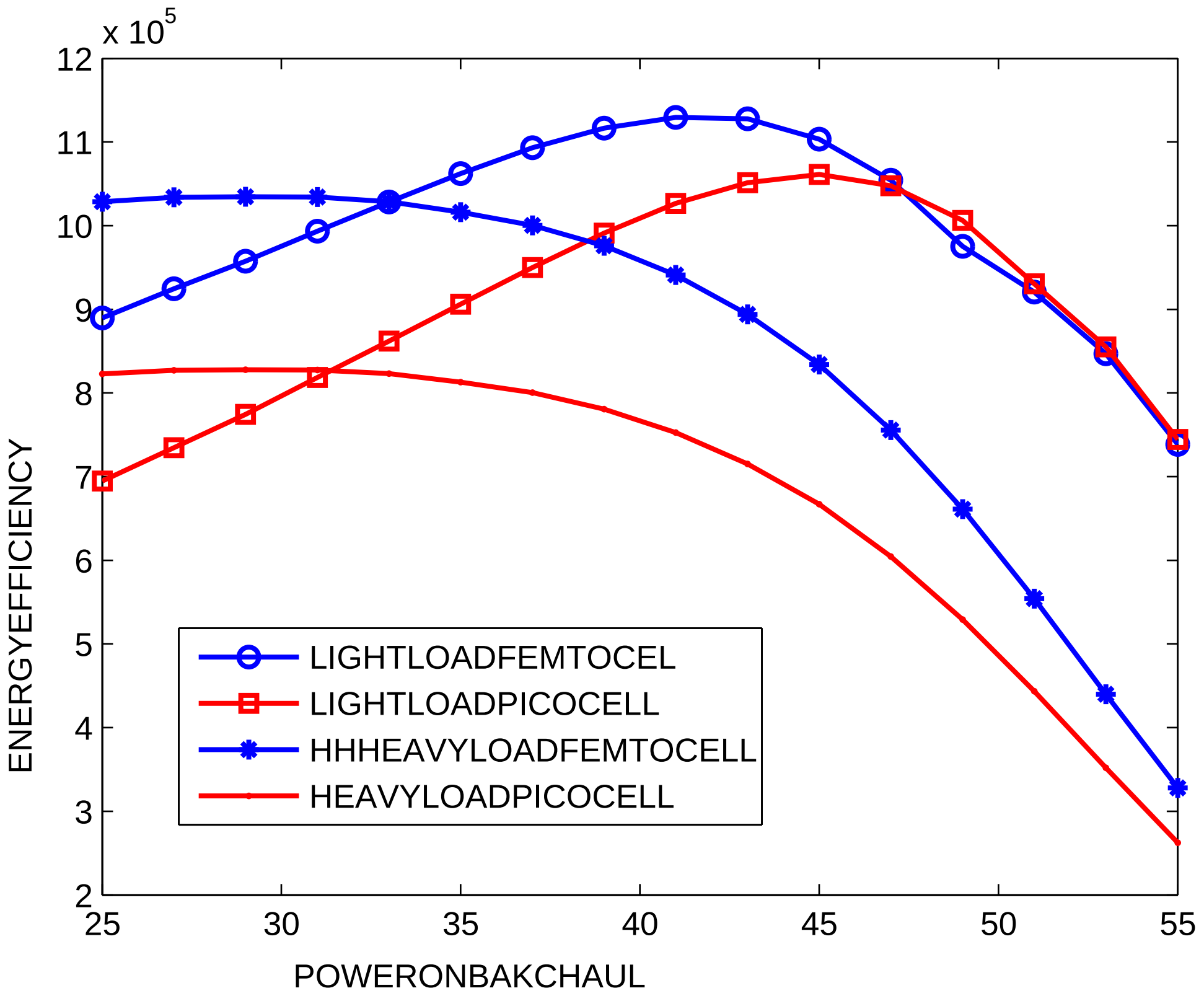

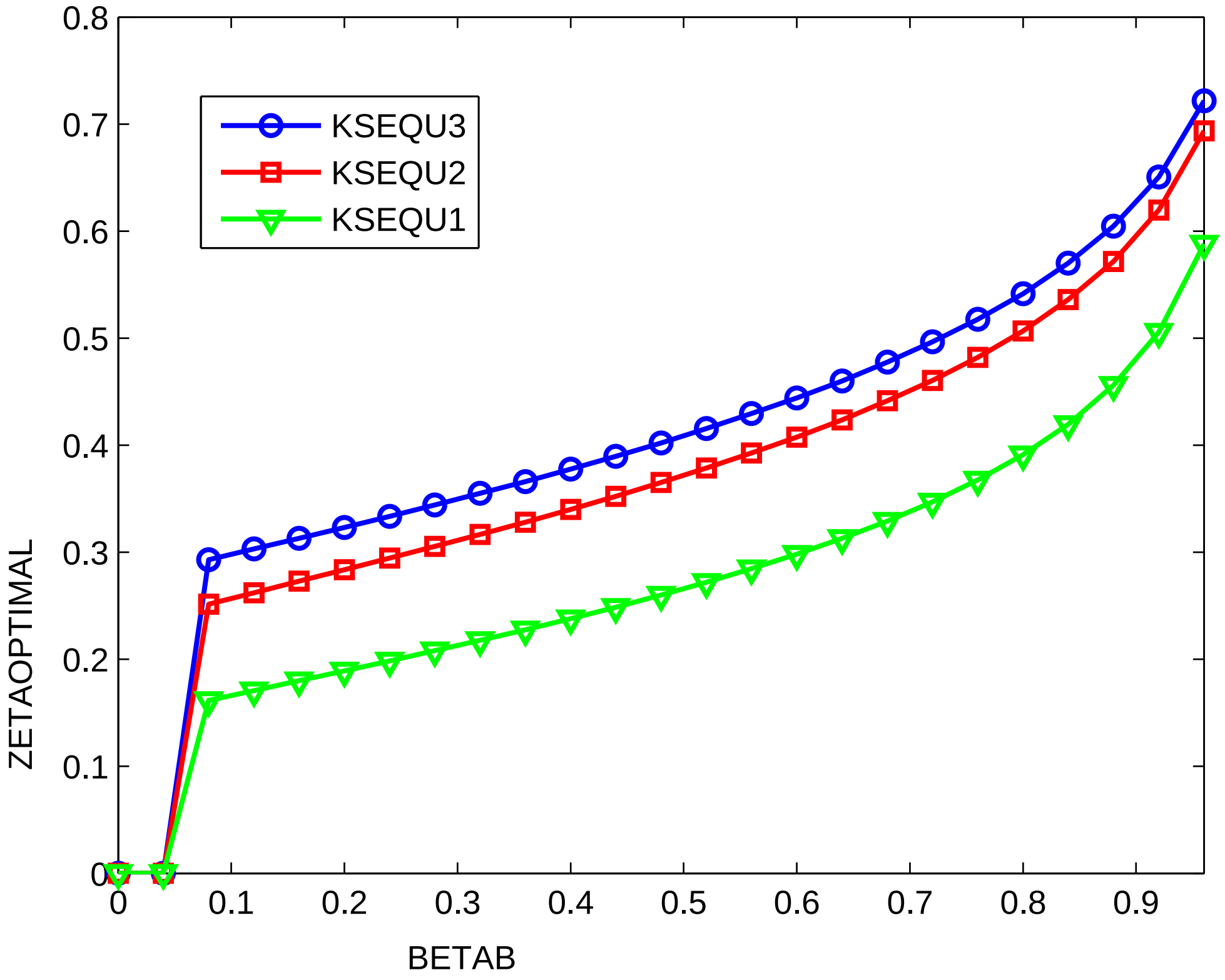

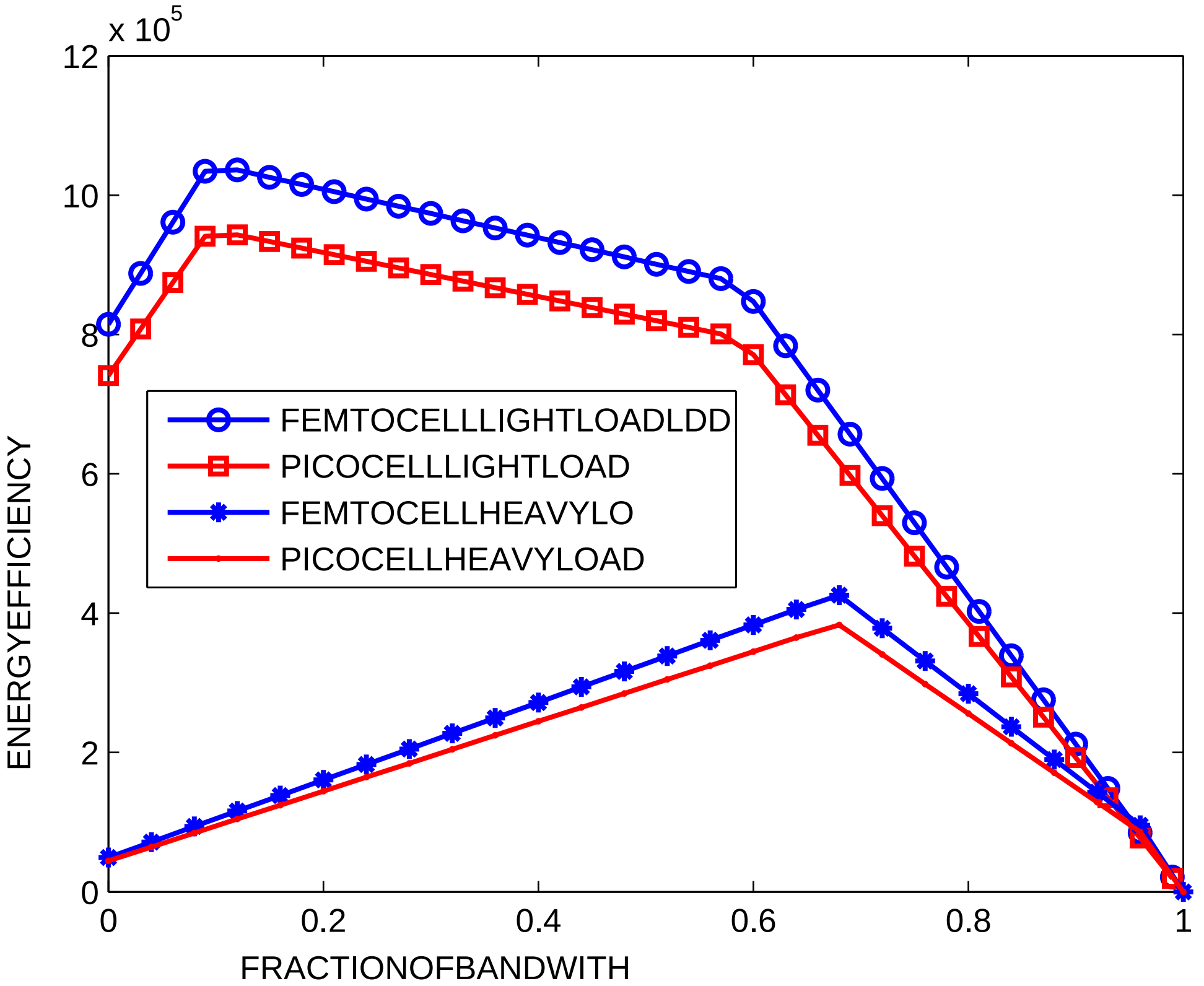